\documentclass[11pt,letterpaper]{article}
\pdfoutput=1

\usepackage{graphicx,array}
\usepackage[dvipsnames]{xcolor}
\usepackage{colortbl}
\definecolor{darkblue}{rgb}{0,0.1,0.5}
\definecolor{darkgreen}{rgb}{0,0.5,0.2}
\definecolor{darkred}{RGB}{153,26,0}
\usepackage{ulem}
\usepackage{slashed}
\definecolor{seablue}{rgb}{0,0.2,0.6}
\usepackage{latexsym}
\usepackage{amssymb, amsmath}
\usepackage{slashed}
\usepackage{bm}        % for math
\usepackage[numbers,sort&compress]{natbib}
\usepackage{bm,relsize}
\usepackage{slashed}
\definecolor{viola}{RGB}{134,41,198}
\usepackage{mathrsfs}
\usepackage{hyperref} %Automatically links \label and \ref commands; Always load last
\usepackage{cancel}
\hypersetup{
    colorlinks=true,       % false: boxed links; true: colored links
    linkcolor=darkblue,          % color of internal links
    citecolor=BrickRed,        % color of links to bibliography
    filecolor=magenta,      % color of file links
    urlcolor=MidnightBlue           % color of external links
}
\usepackage[all]{hypcap} %Link navagates to top of figure instead of caption (below fig)
\usepackage{subcaption}

\usepackage[font=small,labelfont=bf,labelsep=period]{caption}

\newcommand{\be}{\begin{equation}}
\newcommand{\ee}{\end{equation}}

\newcommand{\brho}{\rho}%Background densities
\newcommand{\bvarphi}{\varphi}%Background field
\newcommand{\cH}{\mathcal{H}}

\newcommand{\DM}{\chi}
\newcommand{\DE}{\varphi}

\newcommand{\lm}{{\cal M}}

\newcommand{\ltpsi}{\Phi}

\newlength{\gridwidth} 

\newcommand{\gridmatrix}[4]{
\begingroup
\settowidth{\gridwidth}{\ensuremath{#4}}
\arrayrulecolor{gray!60}
\left[
\begin{array}{c|>{\centering\arraybackslash}m{\gridwidth}}
\rule[-0.25cm]{0pt}{0.6cm}\ensuremath{#1} & \ensuremath{#2} \\ \hline
\rule[-0.5cm]{0pt}{1.2cm}\ensuremath{#3} & \ensuremath{#4} \\
\end{array}
\right]
\endgroup
}

\begin{document}

%%%%%%%%%%%%%%%%%%%%%%%%%%%%%%%%%%%%%%%%%%%%%%%%%%%%%%%%%%%%%%%%%%%%%%%%%%
\begin{flushright}

\end{flushright}
\vspace{.6cm}
\begin{center}
{\LARGE \bf 
Mimicking Phantom Dark Energy \\with Evolving Dark Matter Mass
}\\
\bigskip\vspace{1cm}
{
Lorenzo La Penna$^a$, \large Alessio Notari$^{b}$, Michele Redi$^c$, 
}
\\[7mm]
 {\it \small
 $^{a}$Dipartimento di Fisica e astronomia, University of Florence, Italy\\
$^{b}$Dipartimento di Fisica, Sapienza University of Rome and INFN, Piazzale Aldo Moro 2, I-00185, Italy\\
$^{c}$INFN Sezione di Firenze, Via G. Sansone 1, I-50019 Sesto Fiorentino, Italy\\
 }
\end{center}

\vspace{.2cm}

\centerline{\bf Abstract} 
\begin{quote}
We present a general method to reproduce a given cosmological background through energy exchange between dark energy (DE) and dark matter (DM). This can be simply realized with a standard quintessence scalar field that controls the DM mass. 
In particular a background with phantom crossing can be effectively realized without introducing ghosts or other pathologies. For example one can reproduce exactly the background that gives the best fit to the recent DESI+CMB+DESY5 data, within the Chevallier-Polarski-Linder (CPL) parametrization of DE. Although the background evolution is identical, the perturbations differ, leading to modified growth of structures. If the DM mass varies at late times, early-time observables are not modified and can reproduce the main predictions of the target model, but late-time observables are affected. We discuss in particular the effects on the matter power spectrum, CMB lensing and ISW effect. 
When reproducing the best fit CPL background model, this scenario generically predicts $\mathcal{O}(10\%)$ deviations in such observables. However, for suitable choices of parameters,  effects on the matter power spectrum can be smaller, motivating a detailed study. In general, energy exchange between DE and DM generates a mismatch between the matter power spectrum and the gravitational potential amplitudes compared to the decoupled case, that can lead to deviations  observable in future experiments.

\end{quote}

\vfill
\noindent\line(1,0){188}
{\scriptsize{ \\ E-mail:\texttt{  \href{mailto:lorenzo.lapenna@unifi.it}{lorenzo.lapenna@unifi.it}, 
\href{alessio.notari@uniroma1.it}{alessio.notari@uniroma1.it},
\href{mailto:michele.redi@fi.infn.it}{michele.redi@fi.infn.it}}}}
\newpage

\newpage

\tableofcontents

\section{Introduction}
The recent observations of the expansion of the universe using Baryon Acoustic Oscillations (BAO) by DESI~\cite{DESI:2024mwx, DESI_II:2025} when combined with CMB~\cite{Planck:2019nip} and Supernovae~\cite{DES:2024tys} have sparked renewed interest in the possibility that the acceleration of the universe might not driven by a cosmological constant but that dark energy (DE) might evolve with time.
Even tough this scenario looks preposterous compared to the simplicity and elegance of a cosmological constant, 
the discovery of dynamical DE  would be groundbreaking and it deserves then to be seriously scrutinized. While in the past the main motivation for DE  was the desire to explain the smallness of the cosmological constant and the coincidence problem, recent work is experimentally driven and has motivated a fresh look into the subject.

A plethora of DE models have been considered to explain the data in the literature. In particular, 
quintessence scenarios are a reasonable possibility that can currently fit the data better
than $\Lambda$CDM~\cite{Notari_2024,DESI_Lodha_II, Freeman_Axion_DE} within a theoretically consistent framework, that in the simplest cases amounts 
to adding a single scalar degree a freedom. It is however fair to say that simple quintessence models cannot achieve a fit as good as the one obtained using a phenomenological  parametrization that allows for {\it phantom} behavior, such as the Chevallier-Polarski-Linder (CPL)~\cite{Chevallier:2000qy, Linder:2002et} parametrization where DE is described by an equation of state $w=w_0+w_a(1-a)$. The cosmological model based on the CPL parametrization of DE is also known as $w_0w_a$CDM. Within the CPL parameterization the best fit requires $w_0+w_a<-1$, signaling a phantom behavior at early times (at redshifts $z\gtrsim 0.5$).  This cannot be reproduced in pure quintessence models, where $w> -1$, and generically leads to pathological behavior, that needs to be cured non-trivially (see e.g.~\cite{Creminelli_2009}). 

In this work we consider a relatively simple approach that can reproduce exactly the phantom regime as in the CPL best fit, without phantom pathologies.  
The basic observation is that in this case, while the equation of state of DE becomes phantom, the total equation of state for the sum of DM and DE satisfies $w_{\rm tot}>-1$. This opens the way to the possibility that the background behavior could be reproduced if DM and DE are part of the same sector.  In the most minimal setup, known as ``unified DM", this could be realized through a single perfect fluid with the required equation of state \cite{PhysRevD.66.043507}. 
This possibility is minimal from the point of view of degrees of freedom -- indeed it adds no degrees of freedom to $\Lambda$CDM --
however it endows DM perturbations with non zero  speed of sound, that would be excluded~\cite{Zaldarriaga_UDM}. Ways out  were studied in~\cite{Bertacca:2008uf, kou2025unifieddarkfluidnull}. 

In this work we effectively realize the phantom behavior, by using perhaps the most minimal possibility of an energy exchange between DM and DE (see \cite{Amendola:1999er, Wang:2016lxa, Fifth_force_Diego_I, Giare:2024smz, Khoury:2025txd, Petri:2025swg, Chen:2025ywv, Silva:2025hxw} for related works and \cite{Craig:2024tky} for a similar idea using neutrinos), that is a DM mass scanned by a quintessence scalar field $\DE$ during its cosmological evolution.  The energy density of such an interacting DM component, that we call $\DM$, at the background level evolves as
\begin{equation}
\rho_{\DM}(a)= \frac{m(\DE(a))}{m_*} \frac{\rho_{\rm DM}^0}{a^3} \,,
\label{eq:DMab}
\end{equation}
where $\rho_{\rm DM}^0$ is the present-day value of the DM energy density of the reference model, $a$  the scale factor and $m_*$ a scale that we will fix to match the DM density at a certain time. Eq. (\ref{eq:DMab})  implies that energy is exchanged between DM and DE so that the energy momentum tensors of DM and DE are not separately conserved,
\begin{equation}\label{source conservation}
\nabla_{\mu} T^{\mu \nu}_{(\DM)} =-\nabla_{\mu} T^{\mu \nu}_{(\DE)}= J^{\nu} \, ,
\end{equation}
with $J^{\nu}$ being the four-vector current acting as a source for both $\DM$ and $\DE$, whose  explicit expression will be derived in the next section. 

The microscopic nature of DM is not very important for our discussion but it is simple to construct explicit models that reproduce the dynamics above.
As long as the number changing interactions are negligible, i.e. $n_\DM \sim a^{-3}$, the evolution of DM energy density is obtained by changing the DM mass. 
For example we can consider DM made of fermions with mass that depends on the quintessence field, 
\begin{equation}
\mathcal{L}= \frac 1 2 (\partial \DE)^2 - V(\DE) +\bar{\chi}(i\slashed{\partial}-m(\DE)) \chi \, ,
\label{eq:example}
\end{equation}
or alternatively DM could be a scalar field $\chi$, with fast oscillations around the minimum, whose mass is modulated by $\DE$, via a $m(\DE)^2 \chi^2$ interaction. 

As we will show, the function $m(\DE) $ and the potential $V(\DE)$ can be chosen to reproduce a given equation of state and in particular phantom crossing as in the best fit CPL background. 
Many observables such as BAO (with the sound horizon fixed at early times) or luminosity distances for Supernovae only depend on the background and are therefore guaranteed to be precisely reproduced in our scenario.  Perturbations however behave non trivially. 
In particular, since  DM and DE are not secluded, their perturbations mix with each other and some residual effects remain that could be observable, for example in the matter power spectrum and in late time effects on the CMB. A crucial property is that the interacting DM, $\DM$, has no speed of sound, avoiding the very strong constraints on unified DM. The evolution is however rather  complex and it will be one of the goals of this work to study the modifications from ordinary quintessence models.  

The paper is organized as follows. In section~\ref{sec:bck} we describe the general algorithm to determine $m(\DE)$ and $V(\DE) $ to reproduce a
given background.   In section~\ref{sec: Pert} we  discuss perturbations, deriving the relevant formulas. 
In section \ref{sec:CPL} we apply the framework to the CPL scenario, focusing on the best fit of DESI data that corresponds to an isolated DE component with phanton equation of state. 
To study perturbations we developed a \texttt{MATHEMATICA\textsuperscript{\textregistered}} (Wolfram Research, Inc.) numerical code that allows to easily include interactions between DM and DE. For phenomenological applications we focus on late-time modifications so that the evolution is practically as isolated species till well after recombination. We find that some deviations from standard evolution are present, in particular studying in detail the matter power spectrum, CMB lensing and the Integrated Sachs-Wolfe effect. We summarize the results in \ref{sec:conclusions}. We discuss initial condition and review the field-fluid correspondence in the appendices.

\section{Evolving Dark Matter}
\label{sec:bck}

In this section we describe the general procedure to reproduce a given background expansion rate, allowing for  energy exchange between DM and DE. We follow in particular~\cite{Fifth_force_Diego_I} where the same setup was considered, for a different purpose. 
While the construction is general we will focus on the CPL parametrization used to fit recent DESI BAO with CMB and DES5Y Supernovae, where naively the DE equation of state  crosses the phantom divide ($w=-1$)
at redshift $z\sim 1$, while in our case we show that the sum of DE and DM respects $w_{\rm tot}>-1$ throughout.
Let us note that that other attempts recently made to fit such datasets with non-minimally coupled scalar fields are actually closely related to our approach, since in the Einstein frame such models reduce to standard gravity, with DE being a scalar field that modulates the mass of matter particles (and therefore also DM particles). A crucial difference, however, is that due to the universal nature of gravitational couplings, such models also modulate the mass of {\it any} particle, including visible matter, and therefore they are much more strongly constrained~\cite{Ye:2024ywg, Adam:2025kve, Wolf:2024stt}.

\subsection{Background}

Our starting point is a ``target'' background produced by isolated fluids DM, DE, baryons and radiation.  
The total energy and pressure densities are given by (we assume no spatial curvature)
\begin{equation}
\begin{split}
\frac{\rho_{\rm tot}(a)}{\rho_c}&=\Omega_{\rm DE}f_{\rm DE}(a)+\frac{\Omega_{\rm DM}}{a^3}+\frac{\Omega_{\rm b}}{a^3}+\frac{\Omega_{\rm rad}}{a^4} \, ,\\
\frac{p_{\rm tot}(a)}{\rho_c}&=w_{\rm DE} \Omega_{\rm DE}f_{\rm DE}(a)+\frac{\Omega_{\rm rad}}{3 a^4} \, ,
\end{split}
\label{eq:background}
\end{equation}
where $\rho_{c,0}=3 M_p^2 H_0^2$ is the critical density, $M_p$ is the reduced Planck mass, $H_0$ is the Hubble rate today and $\Omega_I$ are the abundances at present time.
%, and DE and DM are independent fluids.
The evolution of each species is determined by the conservation of the individual energy momentum tensor. In particular for DE one finds,
\begin{equation}
f_{\rm DE}(a)= \exp\left[\int_a^1 3(1+w_{\rm DE}) \frac {da'}{a'}\right] \, .
\end{equation}
The conformal Hubble rate is determined as,
\begin{equation}
{\cal H}^2 = \frac {a^2 \rho_{\rm tot}(a)}{3 M_{\rm p}^2} \, .
\end{equation}

Let us now consider a dark sector where the DM, $\DM$, has a mass that depends on a quintessence field $\DE$, such as the one described in eq.~(\ref{eq:example}). For the sake of this discussion we assume the relevant dynamics to take place well after recombination. One way to think about this system is that the interaction changes the energy of DM but it does not change the number density so that
\begin{equation}
\frac{\rho_{\chi}(a)}{\rho_{c,0}} = \frac{m(\DE(a))}{m_*}\frac {\Omega_{\rm DM}}{a^3} \, ,
\label{eq:rhoDM}
\end{equation}
where $\Omega_{\rm DM}$ is the present-day fraction of a corresponding constant mass DM that evolves in the standard way
\begin{equation}\label{eq:DM isolated evo}
\frac{\rho_{\rm DM}}{\rho_{c,0}} = \frac{\Omega_{\rm DM}}{a^3}\, .
\end{equation}

On the other hand the pressure of DM remains zero so that DM continues to be cold. As mentioned, this feature is crucial phenomenologically because very strong constraints exist if DM is even relatively warm \cite{Irsic:2023equ, Enzi:2020ieg}.
For DE instead we have,
\begin{equation}
\rho_\DE = \frac {\DE'^2} {2a^2} + V(\DE)\,,~~~~~~~~~~~~~~~p_\DE =\frac {\DE'^2} {2a^2} - V(\DE) \, ,
\label{eq:rhops}
\end{equation}
and $w_\DE = p_\DE/ \brho_\DE$ as usual.
Because the sum of DE and DM remains isolated the total energy momentum tensor of the dark sector is conserved. Using eq. (\ref{eq:rhoDM}) this implies,
\begin{equation}\label{eq:contbck}
\begin{split}
&\brho'_{\DM} + 3 \cH \brho_{\chi} =  \brho_{\chi} \lm \, ,\\
&\brho'_{\DE} + 3 \cH \brho_{\DE} (1+w_{\DE}) = - \brho_{\DM} \lm \, ,
\end{split}
\end{equation}\\
where 
\begin{equation}
\lm = \frac{\partial \log m(\DE)}{\partial \eta} \, .
\end{equation}
Throughout prime denotes derivative with respect to the conformal time $\eta$.
Note that $\lm$ plays a similar role to Hubble expansion. Indeed it can be interpreted as an extra force acting on DM \cite{Fifth_force_Diego_I}.

We can now impose that the energy density of the dark sector equals the target sector with isolated components, and the same for pressure.
This determines $\brho_{\DE}$ and $p_{\DE}$,
\begin{equation}
\begin{split}
\brho_{\DE} &= \brho_{\rm DE} + \rho_{c,0} \frac{\Omega_{\rm DM}}{a^3}\left[ 1 - \frac{m(\varphi(a))}{m_*} \right] \, , \qquad \qquad w_{\DE} = w_{\rm DE} \frac{\brho_{\rm DE}}{\brho_{\DE}} \, .
\end{split}
\label{eq:newDE}
\end{equation}
One can easily check that $\brho_{\DM}$ and $\brho_{\DE}$ satisfy eq.~(\ref{eq:contbck}). The second equation just follows from the fact DM is pressureless and thus the total pressure $p_{\DE}=w_{\DE} \rho_{\DE}$ must be equal to $p_{\rm DE}=w_{\rm DE} \rho_{\rm DE}$.
This equation shows than even if $w_{\rm DE}<-1$ the equation of state of coupled DE can satisfy $w_\DE>-1$, if $\rho_\DE$ is increased compared to the case with separate fluids.

From (\ref{eq:rhops}) we can reconstruct $\DE(a)$ and the potential that reproduces the background with isolated components with arbitrary $w_{\DE}$. 
One finds\\
\begin{equation}
\begin{split}
&\frac {d\DE}{da} =\frac 1{\cal H}\sqrt{\brho_\DE (1+w_\DE) } \, ,
\\
&V(\DE(a))  =\frac{1}{2} \brho_\DE (1-w_\DE) \, .
\end{split}
\label{eq:reconstruction}
\end{equation}
The square root can be solved as long as $w_\DE>-1$, which is just the fact that quintessence scalars cannot cross the phantom line. 
Crucially, however, this condition can be realized even though $w_{\rm DE}$ drops below -1, if the DM  compensates for the missing energy density. 
As we will see this allows to reproduce the CPL background that provides the best fit of the DESI + CMB + Supernovae data.

For any real solution $\DE(a)$ we can invert this function and plug it into the second equation. This allows to reconstruct the potential $V(\DE)$ that produces  the starting background.
We will show explicitly a potential that reproduces a phantom background while $w_\DE>-1$ in section \ref{sec:CPL}. This procedure generalizes the algorithm of \cite{Notari_2024} to reproduce a given equation of state within quintessence models.

\section{Perturbations of Evolving Dark Matter}
\label{sec: Pert}

Having reproduced the same background of isolated DM and DE components with evolving DM we next consider perturbations. This is particularly relevant because any difference with the decoupled target model can only arise once perturbations are considered \cite{PhysRevD.80.123001}. In other words, the dark degeneracy is broken when considering inhomogeneities. As we will see this leads to significant differences in the linear perturbations probed in the CMB and large scale structure observations.\\ 

We will study linear order cosmology with scalar metric fluctuations in conformal Newtonian gauge \cite{Ma:1995ey}. The metric is parametrized by
\be\label{Newtonian metric}
ds^2=a^2\left[ -(1+2\Psi)d\eta^2 + \delta_{ij}(1-2\Phi)dx^i dx^j \right] \, ,
\ee
where $\Psi(\eta, \vec{x}), \Phi(\eta, \vec{x})$ are the two gravitational potentials.\\
For our purpose it will be sufficient to study the evolution at low redshift when radiation is negligible. Moreover $\varphi$ has no anisotropic stress at linear order as shown in appendix \ref{app 1}. As a consequence, even though this assumption is not necessary, the analysis is simplified assuming $\Phi=\Psi$ and considering only DM, DE and baryons throughout.

The equations for the perturbations were derived in~\cite{Fifth_force_Diego_I}.
They can be derived as follows. In the background of $\DE$ the worldline action of $\DM$ massive particles is just,
\be\label{world-line action DM}
S_\DM = -\int d\tau \, m(\DE) \sqrt{-g_{\mu \nu} \frac{dx^\mu}{d\tau} \frac{dx^\nu}{d\tau}} \, ,
\ee
where $\tau$ is the proper time along trajectories.
This implies that $\chi$ sees an effective metric $\tilde{g}_{\mu\nu}(x)=m(\DE(x))^2 g_{\mu\nu}(x)$. This observation carries over to the long distance limit where DM is described by a fluid. Therefore the exact classical equations of motion are just the conservation of the energy-momentum of a  pressureless fluid in the auxiliary metric $\tilde{g}_{\mu\nu}$,
\begin{equation}
\tilde{\nabla}_\mu \tilde{T}^\mu_\nu \equiv \partial_\mu\tilde{T}^\mu_\nu+\tilde{\Gamma}_{\mu\alpha}^\mu \tilde{T}^\alpha_\nu-\tilde{\Gamma}_{\mu\nu}^\alpha \tilde{T}^\mu_\alpha=0\,,~~~~~\tilde{T}_{\mu\nu}=\frac 2 {\sqrt{-\tilde{g}}}\frac {\delta S}{\delta \tilde g^{\mu\nu}(x)} \, ,
\end{equation}
where,
\begin{equation}\label{New Connection vs old}
\tilde{\Gamma}_{\mu\nu}^{\rho}=\Gamma_{\mu\nu}^{\rho}+ \frac{\partial \log m(\varphi)}{\partial \varphi} (\partial_\mu \DE \, \delta^\rho_\nu +\partial_\nu \DE \, \delta^\rho_\mu-\partial^\rho \DE g_{\mu\nu}) \, .
\end{equation}
We can recast these equations in terms of the ordinary energy momentum tensor $T_{\mu\nu}=m^2(\varphi) \tilde{T}_{\mu\nu}$ that enters Einstein's equations.
The equations take form,
\begin{equation}
\nabla_\mu T^\mu_\nu=\frac{\partial \log m(\varphi)}{\partial \varphi} \partial_\nu \DE \, T\, ,
\end{equation}
with $T=T^{\mu}_{\mu}$.
This defines the source appearing in (\ref{source conservation}).  Let us note that this equation is valid to all orders in cosmological perturbations. We consider the DM fluid where $T_{\mu\nu}= \rho \, u_\mu u_\nu$.
At the background level the equation above just reproduces eq.~(\ref{eq:contbck}). At linear order one can show that perturbations in the field can be exactly mapped into fluid variables (see \cite{mainini2008scalar} and appendix \ref{app 1} for details) as follows
\be
\begin{split}
&\frac{\delta \DE}{\bvarphi'} = \frac{\theta_\DE}{k^2} \, ,\\
&\frac{\delta \DE'}{\bvarphi'} = \Psi + \frac{1}{1+w_\DE}\left[ \delta_\DE - \frac{1}{2}\left( \frac{\brho'_\DE}{\brho_\DE} (1-w_\DE)-w'_\DE \right) \frac{\theta_\DE}{k^2} \right] \, ,
\end{split}
\ee
where primes denote derivative with respect to  conformal time, the variable $\delta_I\equiv\delta \rho_I / \brho_I$ is the density contrast of species $I$ and $\theta_I\equiv \partial_i v_{I}^{i}$ is the divergence of the velocity perturbation.
One finds then the following equations of motion for dark matter
\begin{align}
&\delta'_\DM +\theta_\DM - 3\Phi' = \lm \left( \ltpsi + \frac{\delta_\DE}{1+w_\DE} \right) + \left[ \lm'+\lm\left( 2\cH + \lm \frac{R_\DM}{1+w_\DE} \right) \right]\frac{\theta_\DE}{k^2} \label{Continuity pert DM 1} \, , \\
&\theta'_\DM + \left( \cH + \lm \right) \theta_\DM - k^2 \ltpsi = \lm \theta_\DE  \, ,
\label{Continuity pert DM 2}
\end{align} 
where $R_\chi = \brho_\DM / \brho_\DE$. When $\lm = 0$ we recover the ordinary equations for cold DM with $\delta P/\delta \rho=0$ because pressure is zero. Every term containing $\lm$ or $\lm'$ comes from the source arising from the energy transfer.

The equations for the quintessence field are similarly determined 
exploiting the conservation of the total energy momentum tensor of DM and DE.
After some algebra one finds,
\begin{align}
&\delta'_\DE + 3\cH \left( \frac{\delta p_\DE}{\delta \rho_\DE} - w_\DE \right) \delta_\DE + R_\DM \lm\left( \ltpsi + \delta_\DM - \frac{w_\DE}{1+w_\DE} \delta_\DE \right) - 3(1+w_\DE) \Phi' + \nonumber \\
&\hspace{5cm}+\left[ (1+w_\DE)k^2 + R_\DM \left( \lm' + \lm \left( 2\cH + \lm \frac{R_\DM}{1+w_\DE} \right) \right) \right] \frac{\theta_\DE}{k^2} = 0\, ,  \label{eq:DEpert1}\\
&\theta'_\DE + \left( \frac{w'_\DE}{1+w_\DE} +\cH(1-3w_\DE) -R_\DM \lm \frac{w_\DE}{1+w_\DE} \right) \theta_\DE - k^2 \left( \frac{\delta p_\DE}{\brho_\DE (1+w_\DE)} + \ltpsi \right) = 0 \, ,
\label{eq:DEpert2}
\end{align}
where $\delta p_\DE$ is the pressure perturbation of the quintessence field, 
\begin{equation}\label{pressure ad+nad}
\delta p_\DE = \brho_\DE \left[ \delta_\DE + \left( w'_\DE - \frac{\brho'_\DE}{\brho_\DE} (1-w_\DE) \right) \frac{\theta_\DE}{k^2} \right]= \brho_\varphi c_{a,\varphi}^{2} \delta_\varphi + \brho_\varphi(1-c_{a,\varphi}^{2}) \left[  \delta_\varphi - \frac{\brho'_\varphi}{\brho_\varphi} \frac{\theta_\varphi}{k^2} \right] \, .
\end{equation}
In the second step we have introduced the adiabatic speed of sound 
$c_{a,\varphi}^{2} \equiv p'_\varphi/\rho'_\varphi = w_\varphi + w'_\varphi \brho_\varphi/\brho'_\varphi$.
Using the background field equations one finds,
\begin{equation}\label{pressure pert quintessence}
\delta p_\DE = \brho_\DE \left\{ \delta_\DE + \Big[w'_\DE + (1-w_\DE) \Big( 3\cH (1+w_\DE) + R_\DM \lm \Big) \Big] \frac{\theta_\DE}{k^2} \right\} \, .
\end{equation}
Importantly DE has non adiabatic pressure, signaling that it is not 
a barotropic perfect fluid with $p(\rho)$. This implies that fluctuations of the field  are not uniquely related to density fluctuations, having contributions from the velocity perturbation $\theta_\DE$.

Crucially, the sound speed in the comoving reference frame, defined as $c_{s,\DE} \equiv (\delta p_\DE/\delta \rho_\DE)\rvert_{\theta_\DE=0}$, is always unity as in any standard quintessence scenario \cite{Creminelli_2009}. Because of this DE perturbations, unlike DM ones, do not grow on small scales.

Finally the equations for baryons remain the standard ones as no energy flows from the dark sector to the visible one:
\begin{align}
&\delta'_b + \theta_b - 3 \Phi' = 0 \label{Continuity b pert 1} \, , \\ 
&\theta'_b + \cH \theta_b - k^2 \ltpsi = 0 \label{Continuity b pert 2} \, .
\end{align}

To close the system we have the linearized Einstein's equations,
\be\label{Einstein perturbed}
\begin{split}
&k^2 \Phi + 3\cH^2 \ltpsi + 3 \cH \Phi' = -\frac{3 \cH^2}{2\brho_{\rm tot}} \left[ \brho_b \delta_b + \brho_\DM \delta_\DM + \brho_\DE \delta_\DE \right] \hspace{2cm} \text{Poisson}\\
&k^2\Phi'+\cH k^2 \ltpsi = \frac{3 \cH^2}{2\brho_{\rm tot}} \left[ \brho_b \theta_b + \brho_\DM \theta_\DM + \brho_\DE (1+w_\DE) \theta_\DE \right] \hspace{2cm} \text{Momentum}\\
&\Phi''+3\cH \Phi' + \cH^2 \Phi + 2\cH'\Phi = \frac{3 \cH^2}{2 \brho_{\rm tot}} \delta p_\DE \hspace{4.2cm} \text{Trace}
\end{split}
\ee
where for simplicity we have assumed no anisotropic stress\footnote{In the numerical analysis of section \ref{sec:CPL} anisotropic contributions due to radiation (mostly massless neutrinos) has been included following  \cite{Ma:1995ey}.
These contributions are relevant during the early radiation dominated epoch and become negligible when the dark sector takes over.} so that $\Phi=\Psi$.
The system of equations \eqref{Continuity pert DM 1}-\eqref{eq:DEpert2}, \eqref{Continuity b pert 1}, \eqref{Continuity b pert 2} and \eqref{Einstein perturbed} is redundant and we can for example use one combination of Einstein's equations together with the equations for $\DE$, $\DM$ and baryons.\\
One can combine Poisson and Momentum equations to determine the gravitational potential,\\
\begin{equation}\label{gravitational potential exact}
\Phi = -\frac{3 \cH^2}{2 k^2 \brho_{\rm tot}} \left[ \sum_I \brho_I \delta_I + 3\frac{\cH}{k^2} \sum_I \brho_I (1+w_I) \theta_I \right] \, .
\end{equation}
However, for computational convenience, we decided to treat the gravitational potential as a free variable, solving the fluids equations together with the Poisson equation in \eqref{Einstein perturbed}.

Let us note that the equation above are written in conformal time. We will however use as input $w_{\rm DE}(a)$ and $m(a)$.
It is thus convenient to change variables so that,
\begin{equation}
\frac{\partial}{\partial \eta}=a {\cal H}  \frac {\partial}{\partial a} \, .
\end{equation}
This allows us to write all equations in terms of $m(a)$ and $w_{\rm DE}(a)$ without actually deriving the potential as a function of the field. It is always possible to reconstruct the potential as described in the previous section.

\subsection{Subhorizon Expansion}\label{sec: subH}

The evolution of density fluctuations simplifies in the subhorizon limit ($k\gg\cH$). Since the set of differential equations is linear, we can write an equation for the density contrast at the price of raising the order of the differential equations. In particular we can write second order equations for the density contrast of each species.  

Since  expressions are lengthy, let us first focus on DE evolution, that is controlled by the equation 
\begin{align}
&\delta''_\DE + \Big[ \cH - 3 \cH w_\DE - \frac{w'_\DE}{1+w_\DE} - R_\DM \lm \Big] \delta'_\DE + k^2 \delta_\DE + R_\DM \lm \delta'_\DM - \frac{3\cH^2}{2 \rho_{\rm tot}}\rho_b (1+w_\DE) \delta_b - \nonumber \\
& - R_\DM \Big[ \frac{3 \cH^2\rho_\DE (1+w_\DE)}{2\rho_{\rm tot}} - \lm' + \lm \Big( (2 -3w_\DE)\cH + \frac{w'_\DE}{1+w_\DE} - \lm \Big( R_\DM \frac{w_\DE}{1+w_\DE} + 1 \Big) \Big) \Big] \delta_\DM = 0\, ,
\label{second order DE}
\end{align}
discarding terms of  order ${\cal O} (1/k)$.
Crucially the pressure term  $k^2 \delta_\DE$  on the first line (associated to the velocity being 1 in the rest frame) dominates for modes well inside the horizon. As a consequence $\delta_\DE$  oscillates rather than growing. This result allows, not only to conclude that $\DE$ is free from gradient instabilities, but also that it does not cluster efficiently. 

Since $\delta_\DE$ does not grow it can in practice be neglected in the equations for DM and baryons. With this simplification DM and baryons satisfy the second order equations,
\be\label{second order DM b}
\begin{split}
&\delta''_\DM + \left( \cH + \lm \right)\delta'_\DM - \left( \frac{3 \cH^2}{2\brho_{\rm tot}} \brho_\DM + \frac{\lm^2 R_\DM}{1+w_\DE} \right) \delta_\DM - \frac{3 \cH^2}{2\brho_{\rm tot}} \brho_b \delta_b = 0\, ,\\
& \delta''_b + \cH \delta'_b - \frac{3 \cH^2}{2\brho_{\rm tot}}\brho_\DM \delta_\DM - \frac{3 \cH^2}{2\brho_{\rm tot}} \brho_b \delta_b  = 0\, .
\end{split}
\ee
Note that  on sub-horizon scales such equations are  $k$-independent so that
the modification to DM and baryon density fluctuations will simply be an overall rescaling function of time. 
It is also convenient to define the combinations,
\begin{align}
&\delta_m \equiv \frac{1}{\brho_\DM + \brho_b} \left( \brho_\DM \delta_\DM + \brho_b \delta_b \right) \, ,\\
&\delta_{\rm diff} \equiv \delta_\DM - \delta_b \, ,
\label{total matter density}
\end{align}
where the former determines the matter spectrum and the latter a possible shift between matter and baryons perturbations. Approximately one finds 
%\MR{Forse scriverei anche la seconda}
\begin{equation}\label{total matter evolution}
\delta''_m + \left( \cH + \lm \frac{\rho_\DM}{\rho_m} \right) \delta'_m - \left( \frac{3 \cH^2}{2\brho_{\rm tot}} \brho_m + \frac{\lm^2 \brho_{\DM}^{2}}{\rho_m \brho_\DE (1+w_\DE)} \right) \delta_m = 0\, ,
\end{equation}
for the total non-relativistic matter overdensity ($\brho_m = \brho_\DM + \brho_b$), while the shift $\delta_{\rm diff}$ remains small.

From the coefficient of $\delta'_m$ we see that $\lm$ contributes as a friction term. Actually for decreasing mass the sign is negative, so this induces more clustering. On the other hand, the term proportional to $\delta_m$ contains two contributions:  the $\lm^2$ gives faster growth, while the ratio $\rho_m/\rho_{\rm tot}$ decreases the growth.
The conclusion is that the modified growth of the the matter perturbation does not have a definite sign. Indeed in explicit examples presented in the next section the matter power spectrum can be smaller or larger than in the decoupled scenario (where however DE perturbations are not consistently included in the phantom region).

\subsection{Initial Conditions}\label{sec:IC}

To solve the equations for the perturbations we need to specify initial conditions. 
For separate species the curvature perturbations are conserved outside the horizon, defined as
\begin{equation}
\zeta_I \equiv - \Phi - {\cal H} \frac {\delta \rho_I}{{\rho_I}'}=-\Phi+ \frac {\delta_I}{3(1+w_I)} \, .
\label{eq:zetaI}
\end{equation}
Adiabatic initial conditions correspond to,
\begin{equation}\label{eq:ad IC rad}
\zeta_I = \zeta_{\rm tot}\equiv - \Phi- {\cal H} \frac{\delta \rho_{\rm tot}}{\rho_{\rm tot}'}\, .
\end{equation}
In radiation domination, where $\Phi \neq \Psi$, and with separate species, these imply the set of initial conditions given in \cite{Ma:1995ey} where also free-streaming neutrinos are considered as source of anisotropic contribution at early times. 

We can write the general adiabatic solution in the superhorizon limit at late times when no anisotropic contributions are relevant.  Imposing that all the curvature perturbations are equal, using Einstein's equations one finds,
\begin{equation}
\begin{split}
&\delta_I=2\left(\Phi+\frac{\Phi'}{\cal H}\right)\frac{\rho_I'}{3{\cal H}(1+w_{\rm tot}) \rho_I} \, ,  \\
& \theta_I= - \frac{k^2}{3{\cal H}(1+w_{\rm tot})} \delta_{\rm tot} \, ,
\end{split}
\label{eq:adbsol}
\end{equation}
with $w_{\rm tot}=p_{\rm tot}/\rho_{\rm tot}$. The value of the velocity field guarantees that non-adiabatic pressure is zero. Note that the adiabatic solution requires that all velocities are equal so that the system behaves as a single fluid for modes that are outside the horizon. 

In the general case where several fluids contribute to the energy density the gravitational potential satisfies,
\begin{equation}
\Phi''+3{\cal H}(1+c_{a,\rm tot}^2) \Phi'+3{\cal H}^2(c_{a,\rm tot}^2-w_{\rm tot})\Phi=0\, ,
\label{eq:bckphi}
\end{equation}
where
\begin{equation}
c_{a,\rm tot}^2= w_{\rm tot}- \frac {w_{\rm tot}' }{3{\cal H}(1+w_{\rm tot})}\, .
\end{equation}
Importantly the potential is only determined by the background and it is constant when $w_{\rm tot}$ is constant.
This ensures that both in the radiation and matter dominated eras, where $w_{\rm tot}=c_{a,tot}^{2}={\rm const}$, the gravitational potential admits a constant and a decaying solution. The latter quickly becomes irrelevant leaving only the constant solution \cite{Wands_2000}.

One can check that (\ref{eq:adbsol}) are exact solutions of the equations in the superhorizon limit at all times of interest on the solutions of eq. (\ref{eq:bckphi}) even with energy exchange.  We have in particular,
\begin{equation}
\zeta_I=\zeta_{\rm tot}=-\frac 2 {3(1+w_{\rm tot})}\left( \Phi+\frac {\Phi'}{\cal H}\right)- \Phi\, .
\end{equation}
Note that $\zeta_I$ are conserved despite the energy exchange, as shown in appendix \ref{superhorizon}.

This shows explicitly the adiabatic solutions as guaranteed by Weinberg theorem \cite{Weinberg_2003} that holds even when energy exchanges are present between species. 

In general non-adiabatic initial conditions, which we do not study in this work, might exist and need to be properly considered especially when dynamical DE is present. In reference \cite{Ballesteros_2010} it is shown that, under fairly general assumption and in the decoupled case, the adiabatic solution is an attractor thus making the system practically insensitive to DE initial conditions. Numerically, the results in our scenario exhibit poor dependencies on DE initial conditions, strengthening the existence of this attractor also in the interacting case. Further studies are however needed. An analysis of isocurvature, i.e. non-adiabatic initial conditions, can be found in \cite{Fifth_force_Diego_I} even though in this work the class of scalar field theories considered is different from our setup.

\section{Reproducing CPL Best Fit With Evolving Dark Matter}
\label{sec:CPL}

We now apply our framework to reproduce, as a phenomenologically relevant example, the CPL background that gives the best fit of the recent DESI+CMB+Supernova data. Specifically we take,
\begin{equation}
w_{\rm DE}= w_0+ w_a (1-a)\longrightarrow f_{\rm DE}=e^{3 (a-1) w_a} a^{-3 (w_0+w_a+1)} \, ,
\end{equation}
with values given in table \ref{tab:param} extracted from the recent analysis \cite{DES_new}. We will focus on the differences between the decoupled scenario as obtained from the Boltzmann cosmological code \texttt{CLASS} \cite{CLASS_I, CLASS-II} and the interacting DE models\footnote{\texttt{CLASS} uses a prescription based on multiple scalar fields known as PPF (Parametrized-Post-Friedmann) \cite{Fang_2008}, to deal with phantom fluids and extract perturbed quantities. Essentially in this scenario DE perturbations are negligible. }.
 For simplicity we will also neglect neutrino masses and leave a full analysis, including MCMC determinations of parameters to future work.

\begin{figure}[t]
  \centering
  \begin{minipage}[c]{0.45\textwidth}
    \includegraphics[width=\textwidth]{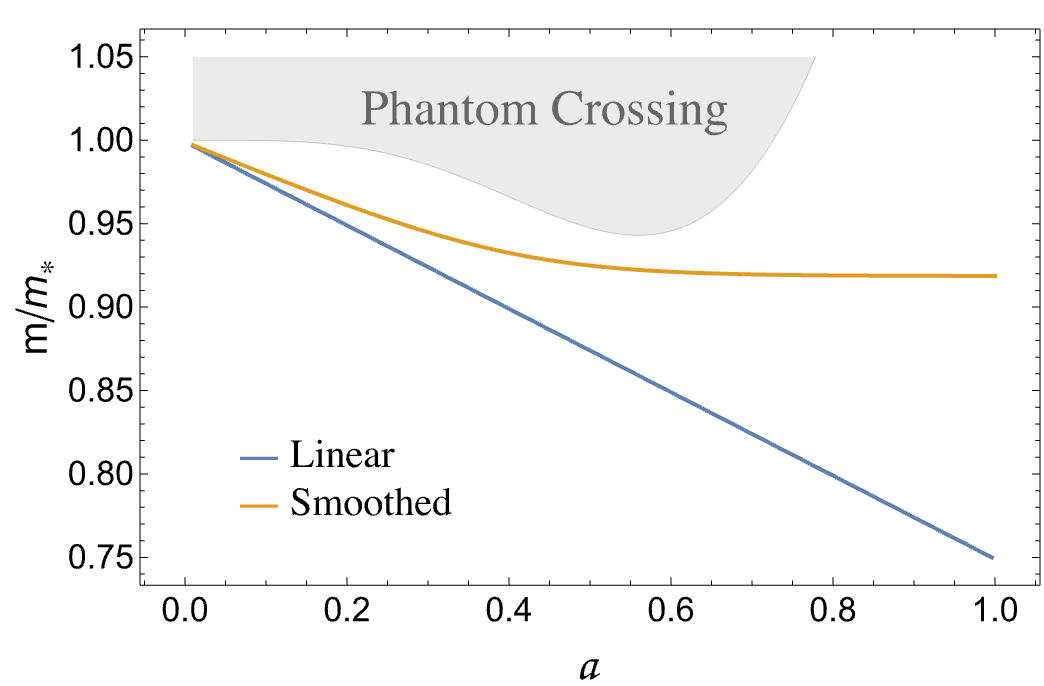}
  \end{minipage}
\hspace{0.02\textwidth}
  \begin{minipage}[c]{0.45\textwidth}
    \includegraphics[width=\textwidth]{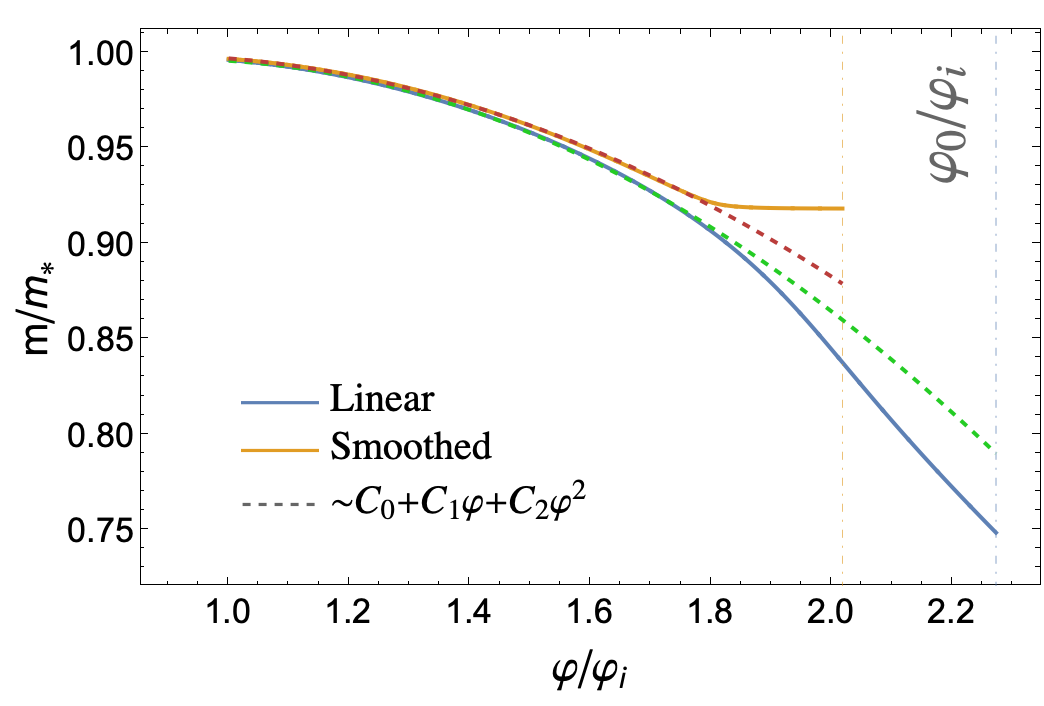}
  \end{minipage}
\caption[DM mass in Linear and Smoothed-Linear models]{\textbf{DM mass in Linear and Smoothed-Linear models.} On the left panel the time dependent mass of the two models. Also shown in gray the region that should not be crossed in order to avoid phantom behavior, for the CPL best fit considered. On the right panel we plot the mass as a function of the quintessence field obtained by solving eq. \eqref{eq:reconstruction}; $\varphi_i$ is a reference field value at $a_i=0.01$.}
\label{CDE mass}
\end{figure}

\begin{table}[h]
\centering
\begin{tabular}{lcccccccc}
\hline\hline
\rowcolor{gray!15}
 & $H_0$ [km/s Mpc$^{-1}$] & $\Omega_m$ & $\Omega_{\rm DM}$ or $\Omega_\DM$ & $w_0$ & $w_a$ & $\alpha$ & $a_s$ & $\Delta$ \\
\hline
$\boldsymbol{w_0w_a {\rm CDM}}$ & 67.47 & 0.313 & 0.264 & -0.803 & -0.72 & - & - & - \\
\textbf{Linear} & / & 0.247 & 0.198 & / & / & 0.25 & $\infty$ & - \\
\textbf{Smoothed-Linear} & / & 0.292 & 0.243 & / & / & 0.2 & 0.4 & 10 \\
\hline\hline
\end{tabular}
\caption[Parameters table in considered models.]{\textbf{Parameters used for the numerical analysis.} A ``/" means that the value is the same as the model above, while ``-" is used if the parameter is not present in the model. The third column is $\Omega_{\rm DM}$ for the decoupled model and $\Omega_{\rm \chi}$ for the linear and smoothed-linear coupled models.}
\label{tab:param}
\end{table}

The decoupled target background features $w_{\rm DE}< -1$ for $a \lesssim 0.6$ that cannot be reproduced in decoupled  quintessence models.
In order to construct a background that avoids phantom crossing, the energy density of DE must be increased compared to the decoupled case, at least until  $a\lesssim 0.6$, so that $w_\DE>-1$ in eq.~(\ref{eq:newDE}). This can be done in several ways. One possibility would be to start at early times with a reduced DM mass and keep the abundance of the CPL best fit today, i.e. a mass that increases with time. This was considered in \cite{Khoury:2025txd}, where however perturbations were not studied in detail. In this case DE redshifts as matter, $\rho\propto 1/a^3$, at early times so it gives a non negligible contribution to the matter density at recombination. While the background can be exactly reproduced, perturbations will be modified at early times. In particular significant modifications to the primary CMB temperature fluctuations at recombination are expected.

Here we consider the simpler possibility that the DM mass instead decreases at late times so that its abundance is equal to the decoupled scenario at early times and smaller today. This corresponds to $m_*\approx m(0)$ in eq. (\ref{eq:rhoDM}). In this way the evolution of perturbations is practically identical to the decoupled scenario at high redshift and changes only at  redshift $z\lesssim {\cal O}(1)$. The only modifications to observables is due to the different evolution of perturbations at low redshift, mostly in the matter power spectrum and in the gravitational potential. For CMB this preserves the successful predictions for the primary fluctuations, leaving only non-trivial secondary effects, in particular lensing and Integrated Sachs-Wolfe (ISW) effects.

\begin{figure}[t]
  \centering
  \begin{minipage}[c]{0.45\textwidth}
    \includegraphics[width=\textwidth]{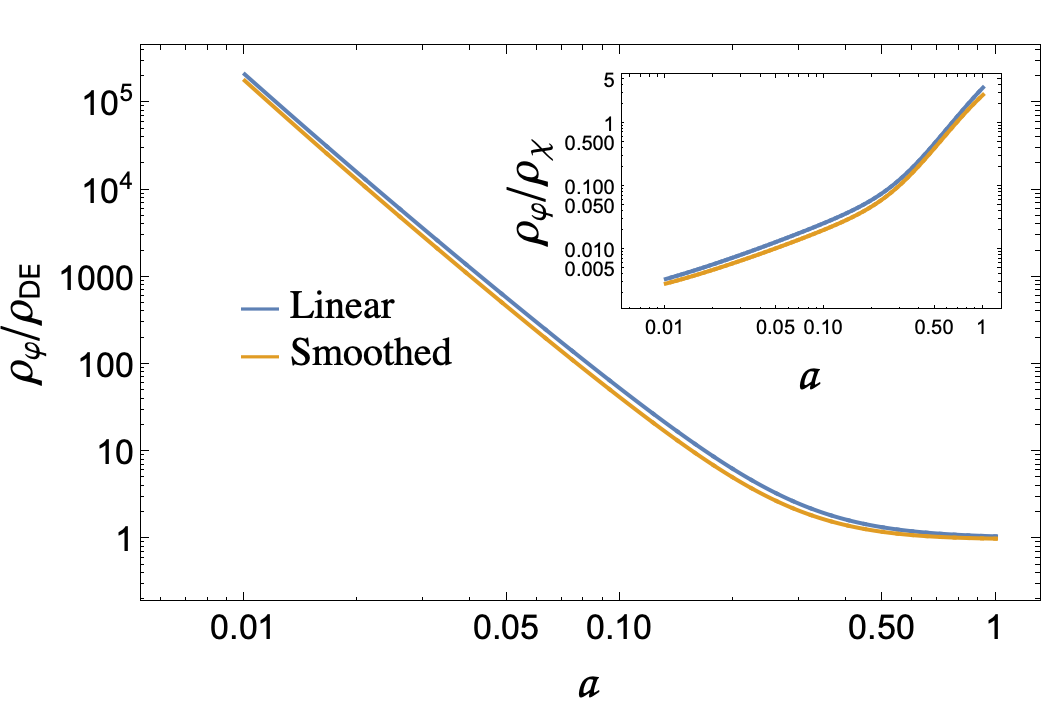}
  \end{minipage}
\hspace{0.02\textwidth}
  \begin{minipage}[c]{0.45\textwidth}
    \includegraphics[width=\textwidth]{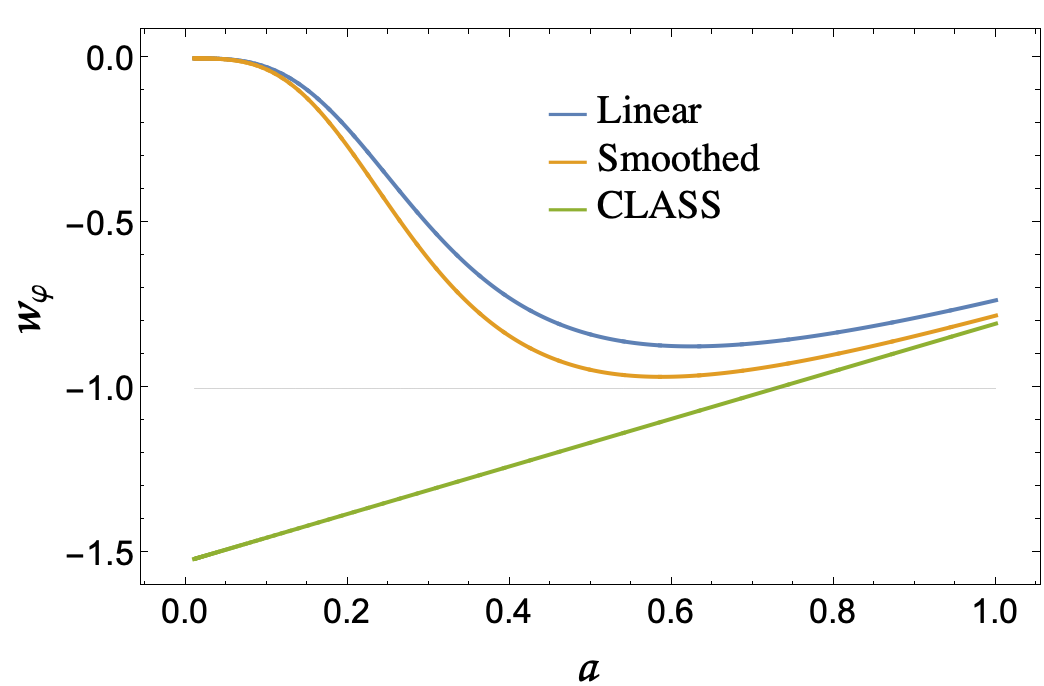}
  \end{minipage}
\caption[DE density and equation of state in CDE models]{\textbf{DE energy density in coupled scenarios.} 
On the left we show the energy density as a  function of the scale factor, normalized to the DE energy density in the CPL uncoupled case. Also shown the ratio of DE over DM density that becomes negligible at early times. On the right we report the equation of state of the coupled DE and of the uncoupled model (``CLASS''), that all give the same best-fit background .}
\label{CDE DE}
\end{figure}

Focusing on the case of decreasing mass we have freedom to choose an arbitrary function $m(a)$ in eq. (\ref{eq:rhoDM}).
The only structural constraint is that $w_\DE>-1$ to solve for the background, demanding an $\cal O$(10\%)  drop of the abundance at late times compared to the CPL bestfit. We have tried several  parametrizations that lead to different behaviors for the perturbations. A large class of models can be captured by the following parametrization,
\begin{equation}
\frac{m(a)}{m_*}= 1-\alpha a+\alpha\big[ S_\Delta(a_s-a) - S_\Delta (a_s) \big] \, ,
\end{equation}
where  $S_\Delta(x) \equiv \frac{1}{\Delta} \log \big[ 1+ e^{-\Delta x} \big] $. 

Within this class of models at early times $m(a)=m_*(1-\alpha\, a)$. We will focus in particular on two representative scenarios. In the first one that we name ``Linear'' we take the mass to be exactly linear in the scale factor that formally corresponds to the limit $a_s\to \infty$ in the parametrization above. 
The second choice that we name ``Smoothed Linear'' instead deviates from linearity at low redshifts. This is motivated by the fact that for $a \gtrsim 0.6$, the equation of state of DE is not phantom and thus one can flatten the decrease of $m(a)$ at such late times, see fig.~\ref{CDE mass}. As we will see this has the effect of reducing deviations from the decoupled case, as derived using  \texttt{CLASS}.

We can easily find the mass and potential that produces this dependence  by solving (\ref{eq:reconstruction}). An analytical solution can be found at early times when DE is still negligible. Assuming matter domination $\rho_\varphi=\rho_c \Omega_{\rm DM}\alpha/a^2$ and $w_\varphi=0$  so that,
\begin{equation}
\begin{split}
m(\varphi)&=m_* \left(1 -\frac {\Omega_{\rm DM}+\Omega_b }{12\Omega_{\rm DM}} \frac {\varphi^2}{M_{\rm p}^2}\right)\, , \\
V(\varphi)&=72 \alpha^3 \rho_c \Omega_{\rm DM}\left(\frac {\Omega_{\rm DM}} {\Omega_{\rm DM}+\Omega_b}\right)^2\frac{M_{\rm p}^4}{\varphi^4}\, .
\end{split}
\end{equation}
Note that the potential diverges at early times. This is due to the fact that, in order to reproduce exactly the background, the field must evolve even at early times. However, from a phenomenological point of view, the tension with $\Lambda$CDM that motivates this work arises mostly from BAO and SNe measures that extend up to $z \lesssim 3$. For higher redshift one can relax the assumption of phantom DE without significant modifications in cosmological observables. This would allow for example to connect the divergent potential with a standard thawing quintessence behavior at early times while reproducing CPL in the phenomenologically relevant redshift range.

The evolution of matter overdensities inside the horizon is controlled by eq.~(\ref{second order DE}) or its simplified version eq.~(\ref{total matter evolution}), so that the solution will depend on the choice of the parameters, $\alpha$, $a_s$ and $\Delta$. Using this freedom we can minimize the difference with the result obtained in the decoupled ``ppf" scenario from CLASS~\cite{Fang_2008}, that we know to be a good fit of the data.
It turns out that in the ``smoothed" benchmark model if  $m(a)$  is chosen to be as smoothly varying as possible, the corrections to the sub-horizon matter power spectrum can be made as small as $1\%$.

\begin{figure}[t]
  \centering
  \begin{minipage}[c]{0.45\textwidth}
    \includegraphics[width=\textwidth]{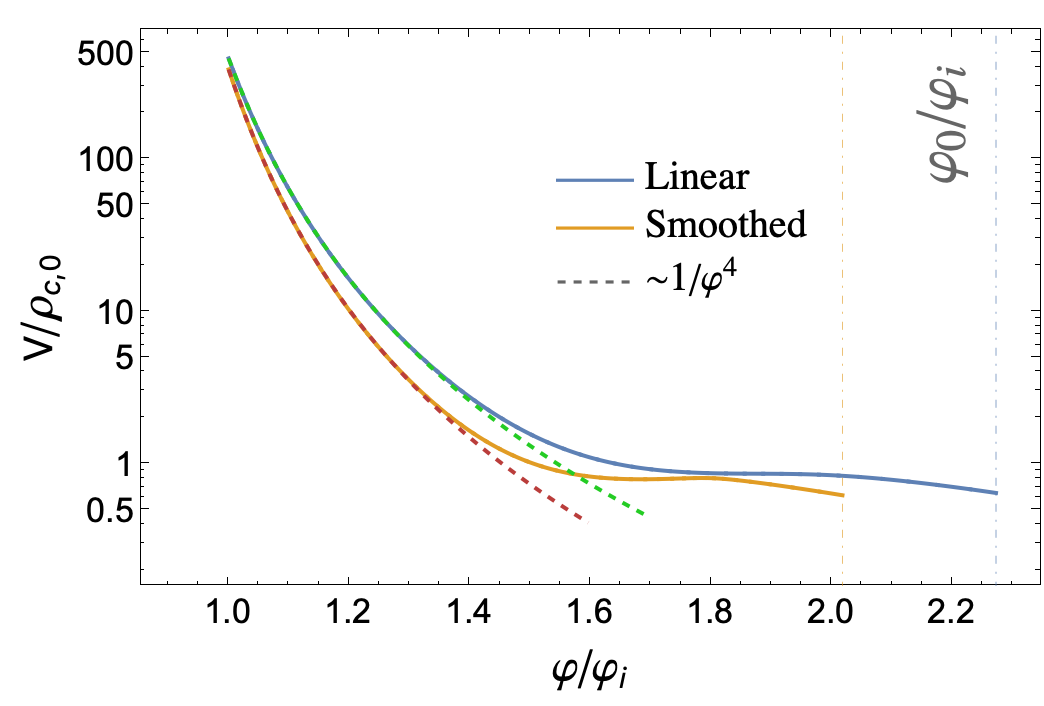}
  \end{minipage}
\hspace{0.02\textwidth}
  \begin{minipage}[c]{0.45\textwidth}
    \includegraphics[width=\textwidth]{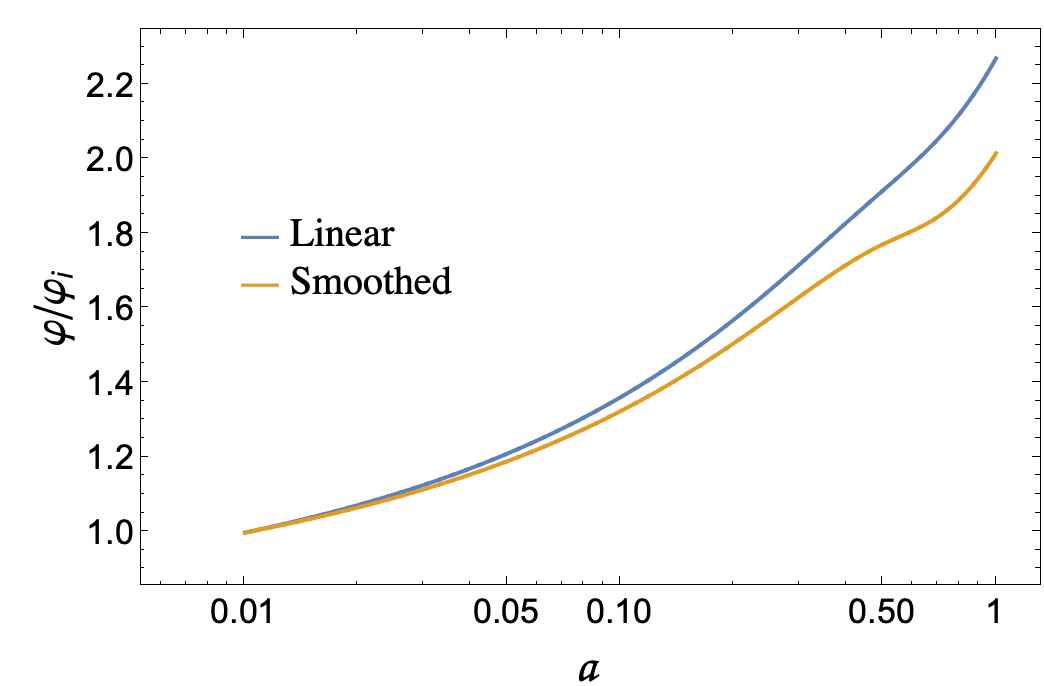}
  \end{minipage}
\caption[CDE Potential and field]{\textbf{Potential of coupled DE models.} On the left the potential  that reproduces the CPL bestfit background, as function of the field over a reference value at $a_i = 0.01$. The evolution of the field as function of scale factor is shown on the right.}
\label{CDE Potential}
\end{figure}

Even if the effects on the matter power spectrum can be made very small, residual effects are however unavoidable on other observables. The effect on lensing signals, such as  CMB lensing,  are  proportional to the gravitational potential $\Phi$. In particular the difference in lensing between our scenario and the decoupled case will be of the order of the offset in $\Phi$ between low and high $k$. Indeed at low $k$ the amplitude of $\Phi$ is fixed by the superhorizon evolution, as seen in section \ref{sec:IC}, which only depends on the background for adiabatic initial conditions. At large $k$ instead $\Phi$ is modified, as can be seen from the Poisson equation,
\begin{equation}\label{Poisson}
k^2 \Phi  \approx -\frac{3 \cH^2}{2 \brho_{tot}} (\brho_{\DM} \delta_{\DM}+\brho_b \delta_b) \, ,
\end{equation}
which simply means that $\Phi$ is essentially determined only by the fraction of matter that can cluster, and not by $\varphi$.
Because $\brho_{\DM}$ and $\delta_{\DM}$ are modified the gravitational potential is modified. Neglecting for simplicity baryons one finds,
\begin{equation}
\frac{\Phi(a)}{\Phi^{\rm CLASS}(a)}\approx \frac{m(a)}{m_*} \frac{\delta_\DM(a)}{\delta_\DM^{\rm CLASS}(a)}\, .
\label{eq:Poissonsimple}
\end{equation}
The gravitational potential is thus determined in terms of the matter overdensity. 

This equation shows that it is not possible to tune at the same time the matter power spectrum and the potential.
For a decreasing mass, as we consider, the potential is  smaller than in the decoupled scenario by a factor $m(a)/m_*$ once ${\delta_\DM(a)}/{\delta_\DM^{\rm CLASS}(a)}$ is tuned to be very close to 1. 

\begin{figure}[t]
  \centering
  \begin{minipage}[c]{0.45\textwidth}
    \includegraphics[width=\textwidth]{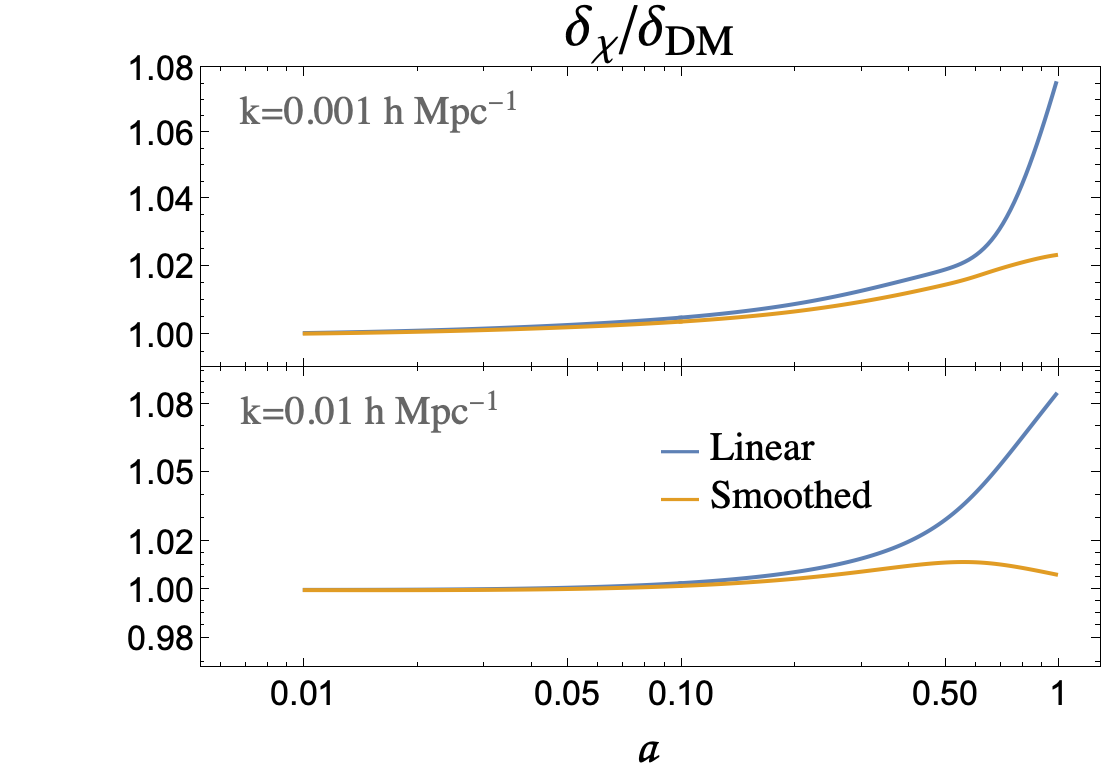}
  \end{minipage}
\hspace{0.02\textwidth}
  \begin{minipage}[c]{0.45\textwidth}
    \includegraphics[width=\textwidth]{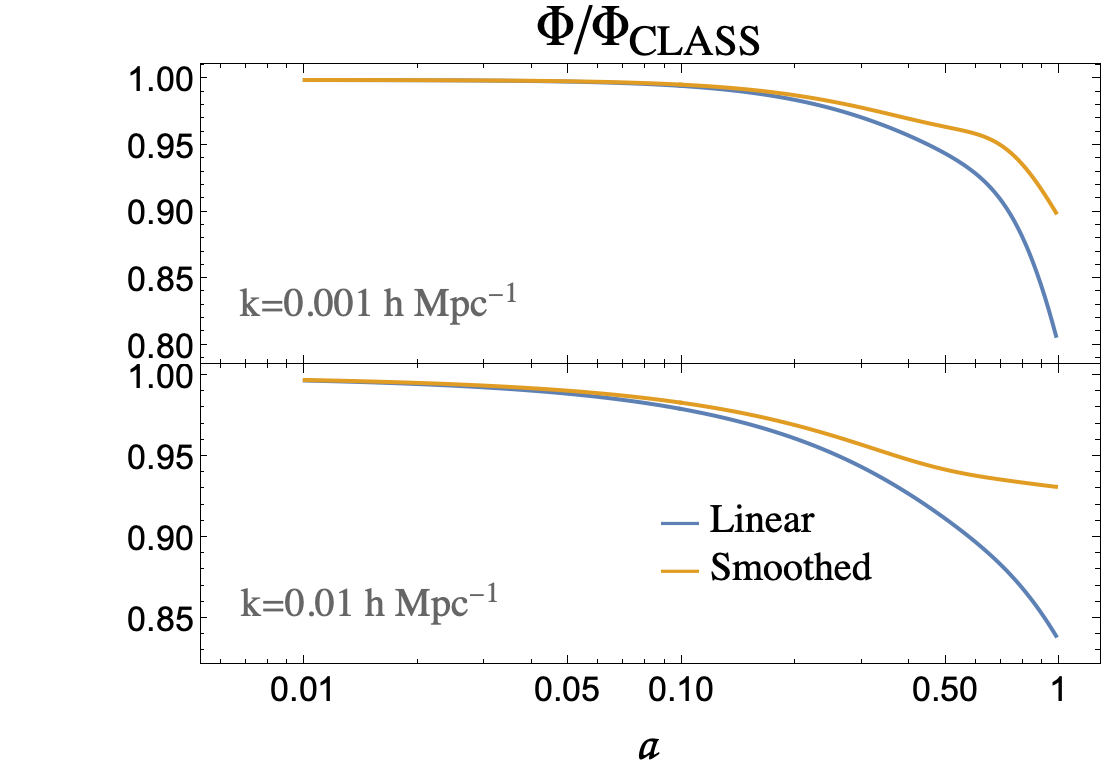}
  \end{minipage}
\caption[DM density contrast and Gravitational potential]
{\textbf{DM overdensity and gravitational potential.} On the left the DM overdensity relative to the same quantity in the uncoupled model as computed in \texttt{CLASS}. On the right the gravitational potential.}
\label{DM density and Phi}
\end{figure}

\subsection{Matter Power Spectrum}

The total matter power spectrum at time $\eta$ is defined  as the two point correlation function of non-relativistic matter density fluctuations\\
\begin{equation}\label{def: matter power spectrum}
\langle \tilde{\delta}_m(\eta, \vec{k}) \tilde{\delta}_m^* (\eta, \vec{k}') \rangle \equiv (2 \pi)^3 \mathcal{P}_m(\eta, k) \delta^{(3)}(\vec{k} - \vec{k}') \, ,
\end{equation}
The tilde over the matter density fluctuations means
\begin{equation}\label{eq: gauge independent matter}
\tilde{\delta}_m \equiv \delta_m -\frac{1}{k^2}\frac{\brho'_m}{\rho_m}\theta_m \, ,
\end{equation}
and corresponds to the  overdensity in the comoving reference frame, where $\theta_m = 0$.
Let us note that the difference between $\delta_m$ and $\tilde{\delta}_m$ is relevant only for modes that re-enter the horizon at late times because the  term proportional to $\theta_m$ is negligible inside the horizon. We recall that we consider massless neutrinos so that the matter power spectrum does not include neutrinos.

\begin{figure}[t]
  \centering
  \begin{minipage}[c]{0.39\textwidth}
    \includegraphics[width=\textwidth]{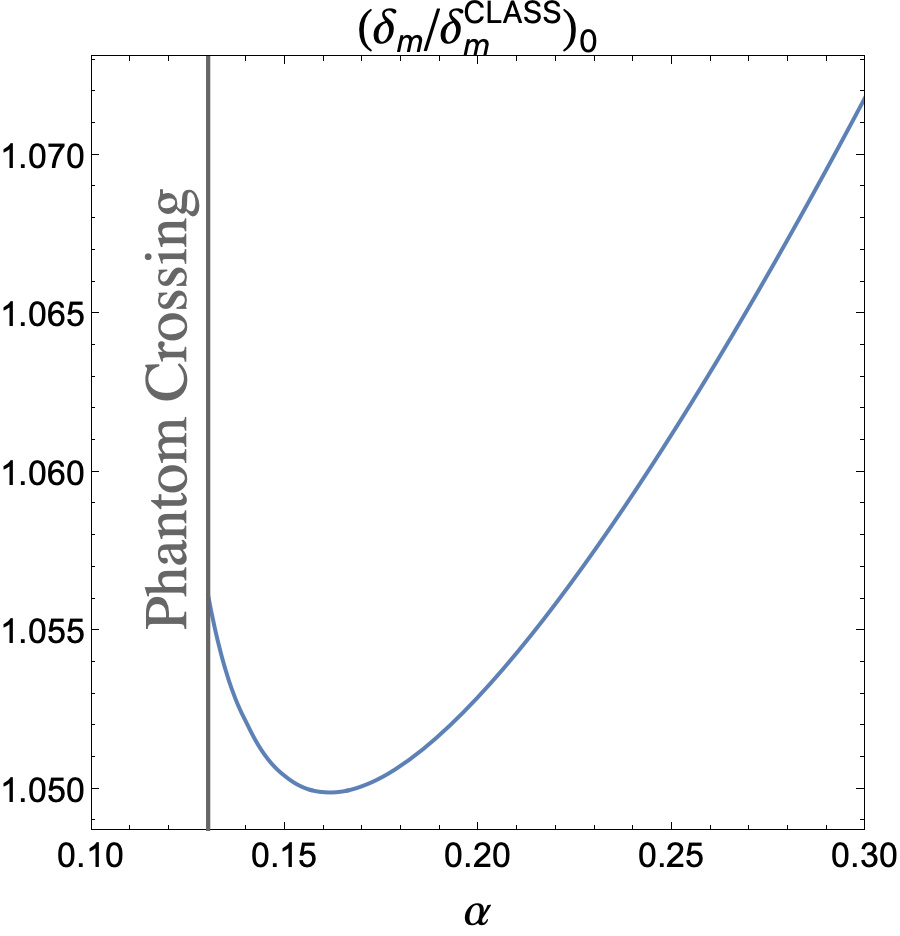}
  \end{minipage}
\hspace{0.02\textwidth}
  \begin{minipage}[c]{0.45\textwidth}
    \includegraphics[width=\textwidth]{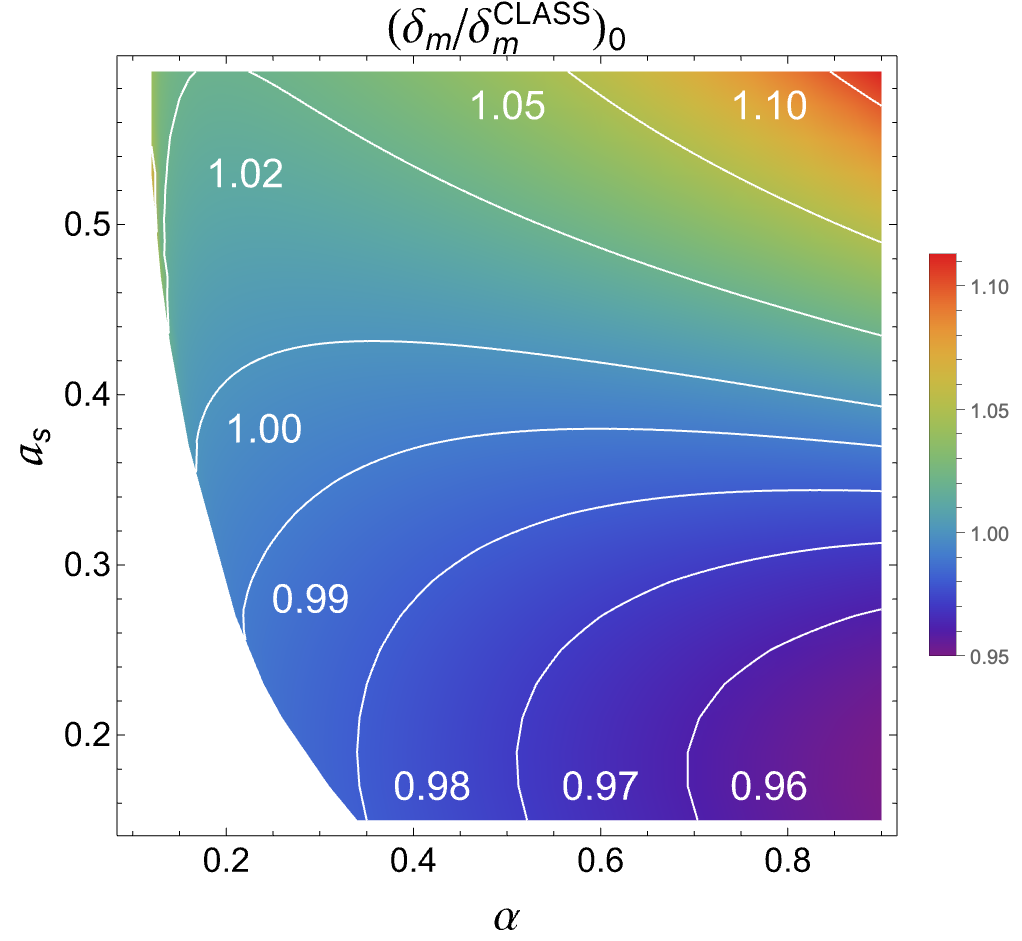}
  \end{minipage}
\caption[Parameter space in the two examinated models]{\textbf{Density contrast of non-relativistic matter.} Left panel shows the increased clustering of matter in the linear model at $z=0$, with respect to the uncoupled model implemented in CLASS. On the right the smoothed scenario is presented where it is possible to realize both reduced and enhanced clustering.}
\label{param space}
\end{figure}

\begin{figure}[t]
\centering
\includegraphics[width=0.6\textwidth]{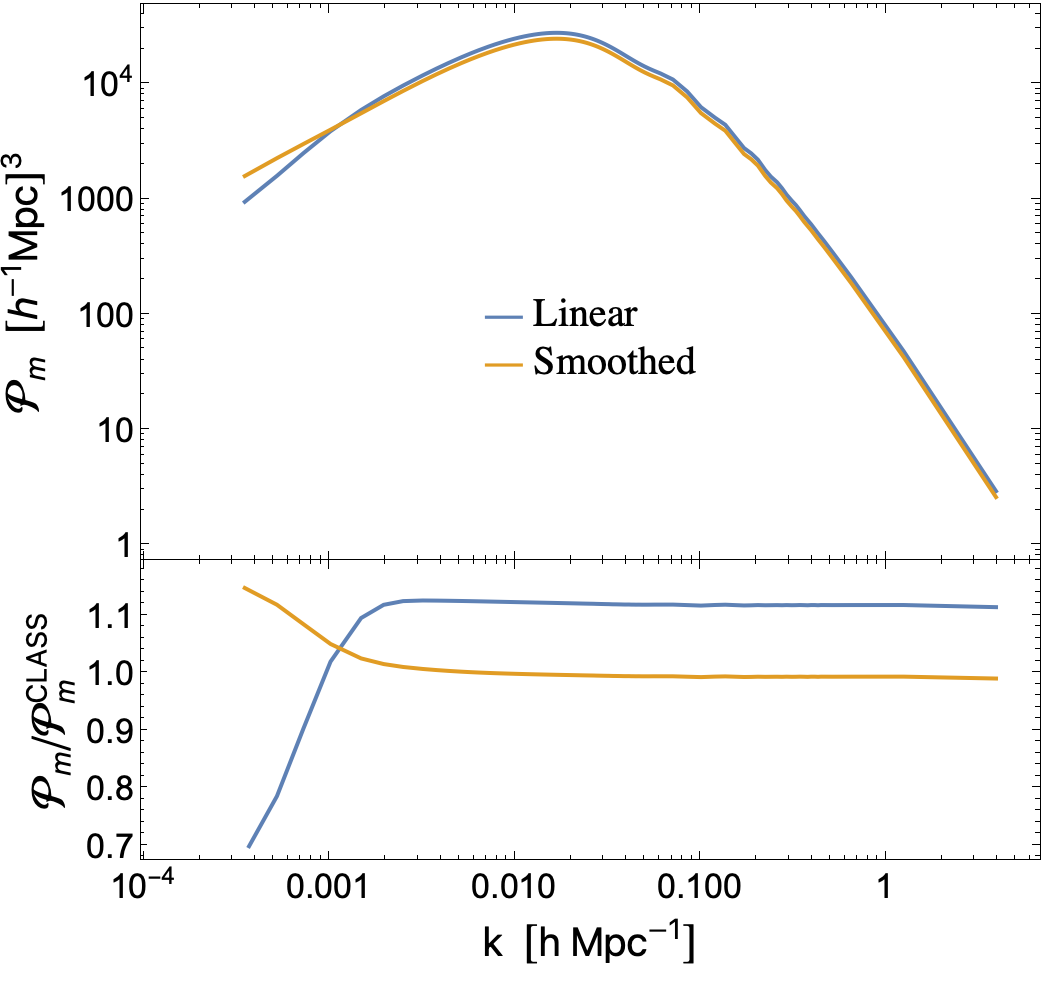}
\caption[Total dimensional matter power spectrum in CDE models]{\textbf{Matter power spectrum at $z=0$.} We show the total matter power spectrum of DM and baryons at present times. On the Lower panel the ratio with the same quantity in the decoupled scenario obtained with \texttt{CLASS} is shown.}
\label{MPS fig}
\end{figure}

Interestingly, the decreasing DM mass does not always lead to an enhanced power spectrum at the present epoch, due to competing corrections, as discussed in section \ref{sec: subH}.
To get some intuition  one can solve numerically the sub-horizon equation~\eqref{total matter evolution}. 
Figure \ref{param space} shows the numerical result obtained by solving \eqref{total matter evolution} compared to the present value obtained in the $w_0 w_a$CDM case with \texttt{CLASS}, while figure \ref{MPS fig} shows the matter power spectrum. As we can see, in the smoothed case both enhanced and suppressed matter clustering are possible, while in the linear model we always find an increase. 

We close this section by noting that the peak of the spectrum corresponds to the wavenumber that enters the horizon at matter-radiation equality, a value that is unchanged in the coupled DE scenario. 

\subsection{Lensing Power Spectrum and Integrated Sachs Wolfe Effect}

As discussed around in the previous section, a structural prediction of our scenario is the mismatch between $\Phi$ and $\delta_m$ at late times. While on large scales $\Phi$ evolves in the usual way at least if initial conditions are adiabatic, after horizon re-entry the evolution is determined by the behavior of the density contrasts $\delta_I$. In particular only $\delta_{\DM}$ grows, while $\delta_{\DE}$ remains roughly constant due to pressure, i.e. gradients in the scalar field equations. This means that the total amount of energy density that clusters is reduced compared to the decoupled case, since the fraction of DM energy density that has been transferred to DE cannot cluster. As a result $\Phi$ is reduced on small scales, as we have seen in eq.~\eqref{Poisson}.\\
This mismatch affects lensing observables. In particular we analyze the lensing potential, that is the effect on CMB photons trajectories modified by the presence of localized distortions in the gravitational potential, see~\cite{LEWIS_2006, Planck2018_Lensing, Carron_2022, Tristram_2024}. Late-time modifications to the gravitational potential also enter the CMB temperature power spectrum through the Integrated Sachs Wolfe effect (ISW) \cite{ISW_original, seraille2024constrainingdarkenergyintegrated, St_lzner_2018, Cabass_2015}, that is the mismatch between energy gain and loss when CMB photons pass through gravitational wells due to the expansion of the universe.

\begin{figure}[t]
\centering
\includegraphics[width=0.6\textwidth]{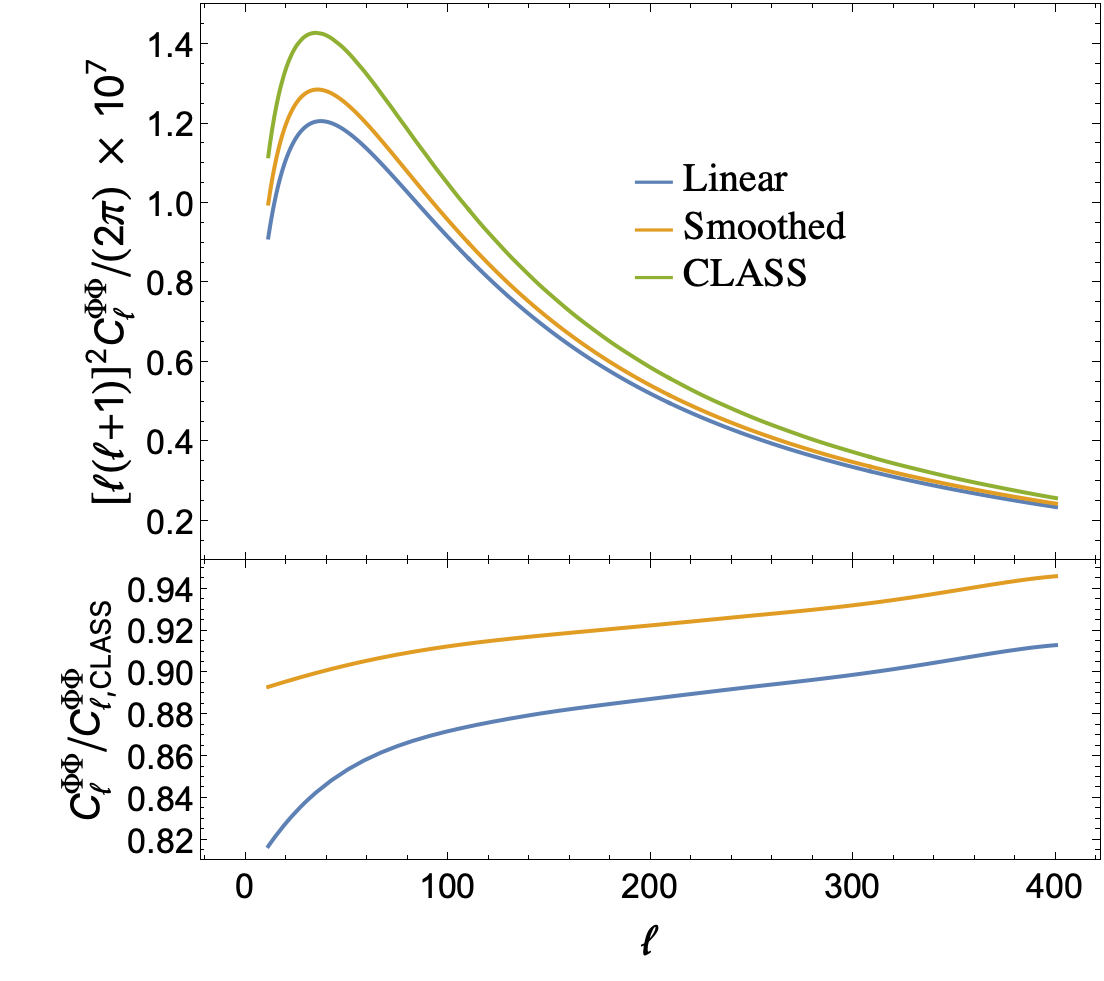}
\caption[Lensing Potential Power Spectrum]{\textbf{Lensing Power Spectrum.} In the upper plot we show results for $w_0w_a$CDM uncoupled model obtained with \texttt{CLASS} and for interacting DE scenarios. In the lower panel we report the ratio of the power spectra. }
\label{Lensing figure}
\end{figure}

Due to the decrease in $\Phi$ we generically expect weak lensing effects to be suppressed. The power spectrum of the lensing potential~\cite{LEWIS_2006} can be computed as
\begin{equation}
C_\ell^{\Phi\Phi}
= 16\pi
\int_0^\infty \frac{dk}{k}\,
\Delta_{\zeta}(k)
\left[
  \int_{\eta{_*}}^{\eta_0} d\eta\,
 T_\Phi(\eta\,,k)\,
  j_\ell(k(\eta_0-\eta)) \,
 \frac{\eta -\eta_*}{(\eta_0 - \eta_*)(\eta_0 - \eta)}
\right]^2\,,
\label{eq: lensing power spectrum}
\end{equation}
where $T_\Phi(\eta\,,k)\equiv\Phi(\eta\, , k)/\zeta_*(k)$ is the transfer function of $\Phi\approx \Psi$ normalized over the initial curvature perturbation, $\eta_*$ the conformal time at the last scattering surface, $j_\ell(x)$ the spherical Bessel function of first type and order $\ell$ and $\zeta(k)$ is the curvature perturbation with power spectrum,
\begin{equation}
\Delta_{\zeta}(k)= A_s \left( \frac{k}{k_\star} \right)^{n_s -1} \,,~~~~~~~~A_s=2.1\times 10^{-9}\,,~~~~~ n_s=0.966\, .
\end{equation}
\begin{figure}[h]
\centering
\includegraphics[width=0.6\textwidth]{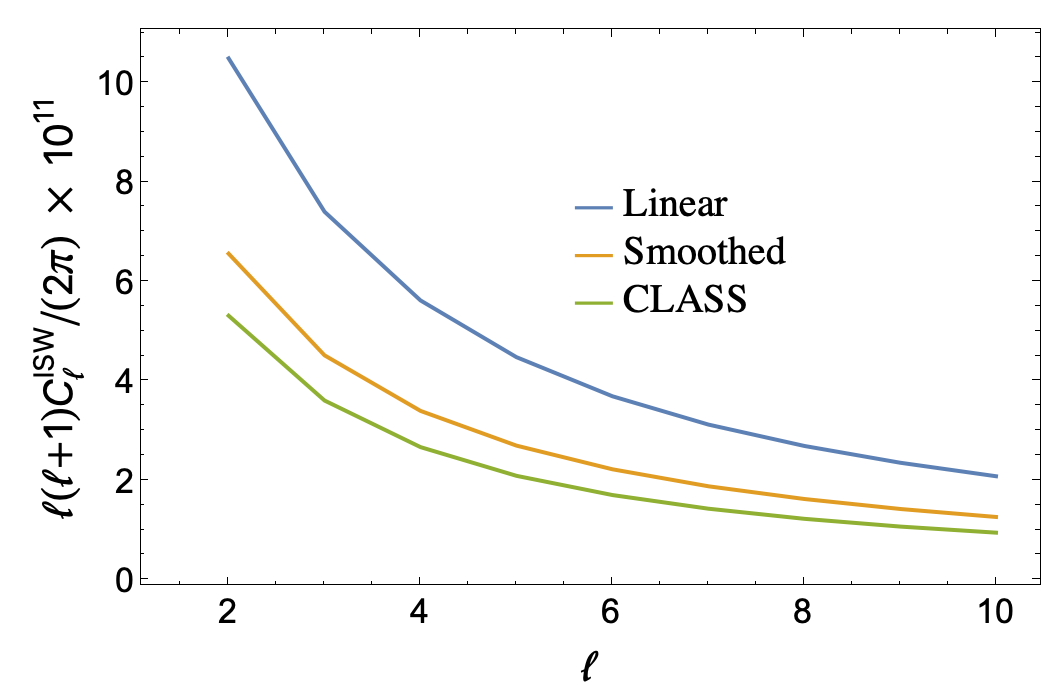}
\caption[ISW]{\textbf{Integrated Sachs-Wolfe effect.} We show the contribution to the late ISW effect in linear and smoothed  scenarios compared to the uncoupled model implemented in \texttt{CLASS}. This is only relevant for small $\ell$ where an enhancement up to ${\cal O}(50\%)$ is found.}
\label{ISW figure}
\end{figure}

Since \eqref{eq: lensing power spectrum} is computationally expensive due to the presence of the double integral, one can adopt the common \textit{Limber} approximation~\cite{Limber_I}\\
\begin{equation}\label{Limber approx}
k \approx \frac{\ell + 1/2}{\eta_0 - \eta} \, ,
\end{equation}
which is fairly accurate for $\ell > 10$. With this approximation the lensing power spectrum can be computed as
\begin{equation}\label{approx lensing power spectrum}
C_{\ell}^{\Phi \Phi} \approx \frac{8 \pi^2}{(\ell+1/2)^3} \int_{\eta_*}^{\eta_0} d\eta\,(\eta_0-\eta) \Delta_\zeta\left(\frac{\ell+1/2}{\eta_0-\eta}\right) \left[ T_\Phi\left( \frac{\ell+1/2}{\eta_0-\eta} \right) \left( \frac{\eta -\eta_*}{(\eta_0 - \eta_*)(\eta_0 - \eta)}\right) \right]^2 \, ,
\end{equation}

Due to  the suppression both at high redshift ($\eta\approx \eta_*$) and at very low redshift ($\eta \approx \eta_0$) given by two different factor in the above expression, the lensing contribution is peaked around $z \approx 3$. This range correspond to late time significant contributions that fall precisely in the times when the energy exchange modifies the linear dynamics. In Fig. \ref{Lensing figure} we show the results for the lensing power spectrum in different models.
From the numerical analysis we find, as expected, a suppression of order $10\%$.
Importantly, this suppression is also present if deviations in the matter power spectrum are reduced to be practically vanishing as can be seen from the smoothed linear parametrization depicted in figure \ref{MPS fig}.

The modified gravitational potential also induces a shift in the late ISW effect, the modification of the temperature perturbation due to the time variation  of the gravitational potential along the line of sight. The contribution to temperature multipoles can be written as ~\cite{Hu_1994}
\begin{equation}
C_\ell^{\rm ISW}
= 16\pi \int_0^\infty \frac{dk}{k}\,
\Delta_{\zeta}(k)\,
\left[
  \int_{\eta_*}^{\eta_0} d\eta\; \frac{\partial T_\Phi(k,\eta)}{\partial \eta}\;
  j_\ell\!\big(k(\eta_0-\eta)\big)
\right]^2 .
\end{equation}

As can be seen from figure \ref{DM density and Phi} the time variation of $\Phi$ is larger than in the decoupled scenario so we expect a larger contribution to ISW.
This is quantified by figure \ref{ISW figure} that shows an enhanced ISW contribution of order $20\%$ to $\mathcal{O}(1)$ depending on the mass choice adopted. Notice that, while the weak lensing affects smaller scales (relevant contributions up to $l\approx500$), the ISW is only relevant for much larger scales ($\ell \lesssim 10$) that do not allow to make appropriate use of the Limber approximation.
This result can be roughly described by an average overall rescaling with respect to the decoupled case.  For this purpose we define a rescaling parameter $A$, whose estimator is given by $\chi^2$ minimization as
\begin{equation}\label{eq:amplitude A}
\hat{A} \equiv \frac{\sum_\ell  C_{\ell} C_{\ell, \, {\rm CLASS}}/\sigma_\ell^2}{\sum_{\ell}  \left(C_{\ell, \, {\rm CLASS}}\right)^2/\sigma_\ell^2}\, ,
\end{equation}
where $\sigma_{\ell}^{2}=2 C_{\ell,\, \rm CLASS}^{2}/(2\ell+1)$ is the cosmic variance, which dominates at low $\ell$.
In the two models under consideration we obtain $\hat{A}_{\rm L} = 0.89,\,0.93$, for the lensing power spectrum, and $\hat{A}_{\rm ISW} = 2.15,\,1.30$ for the ISW respectively for the linear and smoothed case.

While we leave a detailed discussion on CMB temperature power spectrum corrections in such scenarios to future works, we note that it is in principle possible to also tune the parameters of the theory in order to alleviate differences both in the lensing power spectrum and ISW contribution by properly adjusting the gravitational potential given by eq. \eqref{Poisson}. On the other hand in doing so one has to require significant changes in the non-relativistic matter density contrast that will produce deviations in the matter power spectrum. 
A detailed analysis inevitably requires the use of Markov-Chain-Montecarlo to fully explore the parameter space. 

\section*{Conclusions}
\label{sec:conclusions}

In this work, we introduced a general framework to reproduce a given cosmological background at late times, through a rolling scalar field that controls the DM mass. We have shown that a stable realization of phantom DE scenarios can be achieved, avoiding any pathological behavior, including ghosts or other instabilities. This provides a physically consistent realization of parameterizations of phantom DE, such as the $w_0w_a$CDM best fit of the recent DESI + CMB + DESY5 data.  
By construction the coupled DE scenario exactly reproduces the same expansion history of the input background. This is achieved modifying the DM abundance at late times so that the only modifications of the dynamics appears in the perturbations at low redshift. The evolution of perturbations is modified so that the growth of structure and related observables differ from the decoupled case. Since we chose to have modifications only well after recombination, where different species behave practically as perfect fluids, we were able to implement the modified evolution equations in a \texttt{MATHEMATICA} code that allows to easily include energy exchange between DM and DE.

As a case study we applied this framework to the CPL parametrization of DE
that has recently received a lot attention following measurements of BAO by DESI and supernovae. 
On large scales, if adiabatic initial conditions hold -- as we assume -- the system behaves effectively as single fluid, and the evolution remains standard for the total curvature perturbation \(\zeta_{\rm tot}\). Once modes enter the Hubble horizon, several new effects arise, proportional to the logarithmic derivative of the DM mass.  On sub-horizon scales, the correction to the matter power spectrum appears as a scale-independent overall shift, which can be either an enhancement or a suppression depending on the competition among different terms.  
By appropriately tuning the parameters, it is possible to achieve cancellations between these contributions, reducing deviations in the power spectrum to below \(\mathcal{O}(1\%)\) with respect to the decoupled case.  

The effects on lensing observables, and in particular on CMB lensing, are determined by the gravitational potential \(\Phi\), which is fixed by the Poisson equation, once the DM density contrast \(\delta_{\DM}\) and background density \(\rho_\DM\) are specified. Even when \(\delta_\DM\) matches the standard, decoupled case, \(\rho_\DM\) is typically smaller by a factor \(m_0/m_{*}\) if the DM mass today is smaller than at early times. Consequently, in such cases the overall lensing potential is generically suppressed. Focusing on  the DESI + CMB + DESY5 best fit, this corresponds to at least a \(5\%\) suppression of \(\Phi\), i.e. about an \(10\%\) suppression of the CMB lensing deflection power spectrum. This effect is directly related to the degree of ``phantomness'' of the corresponding background trajectory, that is, to the fact that \(1 + w_\DE \simeq -0.05\) at \(z \sim 1\). The lensing constraints could however be weakened by choosing a CPL background still within, for example, the \(1\sigma\) region of the recent DESI + CMB + DESY5 data fit.  

The above considerations motivate a dedicated future study where the model is implemented into a Boltzmann code, to perform a full Markov-Chain-Montecarlo analysis and determine whether there exist regions of parameter space where both the background expansion and the perturbations remain consistent with current data. More generally, beyond the DESI + CMB + DESY5 data, we emphasize that this framework provides a well-defined setting to explore novel effects on perturbations, since the standard relation between \(\delta_\DM\) and the gravitational potential \(\Phi\) is altered inside the horizon. This could, for instance, allow to reconcile measurements based on different observables\footnote{For example, the so-called $\sigma_8$ or $S_8$ tension, i.e. a discrepancy between the high $S_8$ value estimated
by Planck CMB assuming $\Lambda$CDM and the lower value preferred
by cosmic shear measurements~\cite{DiValentino:2020vvd}, which could be naturally addressed by a decreased gravitational potential at late times, though such tensions are alleviated in the most recent data releases~\cite{Kilo-DegreeSurvey:2023gfr,Wright:2025xka,Stolzner:2025htz}.}.  

We believe that this work opens several promising directions for further investigation, with potential relevance for both current and forthcoming cosmological experiments that we are planning to pursue.

\subsubsection*{Acknowledgements}
We thank A. Tesi, D. Redigolo, M. Simonovic, E. Salvioni, G. Ballesteros for useful discussions. The work of MR is
supported by the European Union – Next Generation EU and by the Italian Ministry of University and Research (MUR) via the PRIN 2022 project n. 20228WHTYC. The work of A.N. is partially supported by the grants
PID2022-137268NB-C52 from the Spanish Ministry of
Science and Innovation, Unit of Excellence Maria de
Maeztu 2024-2027 of ICCUB, University of Barcelona (CEX2024-001451-M) and AGAUR 2021 SGR 00872.

\appendix

\section{Initial Conditions and Super Horizon Equations}
\label{superhorizon}

In this appendix we provide further details on the initial conditions discussed in section \ref{sec:IC}.

Let us first consider  the decoupled fluids. From the definition of the curvature perturbations (\ref{eq:zetaI}), using the continuity equation of the fluids, on superhorizon scales one finds for any fluid,
\begin{eqnarray}
\zeta'_I=\frac{(c_{a, I}^2-c^2_{s, I}) \cH}{1+w_I} \tilde{\delta_I} \, , \label{zetadot}
\end{eqnarray}
where $c_{a}^{2}, \, c_{s}^{2}$ are  the adiabatic and comoving speed of sound, $\tilde{\delta}$ is the comoving density contrast given by
\begin{equation}\label{eq:com delta}
\tilde{\delta}_I \equiv \delta_I - \frac{\rho'_I}{\rho_I}\frac{\theta_I}{k^2} \, .
\end{equation}

Therefore, $\zeta_I$ is conserved provided $c_{a,I}^{2}=c_{s,I}^{2}$, i.e. the fluid is barotropic, or $\tilde{\delta}=0$. For a scalar field the former condition is not satisfied so that the curvature perturbation is constant only if the overdensity vanishes in the comoving frame.

The condition $\tilde{\delta}=0$ is always true for a single field system in the superhorizon limit
as this follows from the combination of Poisson and momentum equation in eqs.~(\ref{Einstein perturbed}).

\subsection{Multi-Fluid without Energy Exchange}

Let us now consider a multi-fluid case: 2 fluids, DM ($\DM$) and a scalar field ($\varphi$), with no energy exchange. 

Subtracting eqs.~\eqref{eq:DEpert2} and \eqref{Continuity pert DM 2} with $\lm=0$ and defining $\Delta\theta \equiv \theta_\varphi-\theta_{\DM}$, we obtain,
\begin{eqnarray}
\Delta\theta'=-\cH\Delta\theta+\frac{k^2}{1+w_\DE} \tilde{\delta}_\DE, \qquad 
\tilde{\delta}_\DE\equiv \delta_\DE +3 \cH(1+w_\DE) \frac{\theta_\DE}{k^2} \, .
\label{eqdeltatheta}
\end{eqnarray}
It is convenient to introduce the isocurvature perturbation, 
\begin{eqnarray}
{\cal S}\equiv 3(\zeta_\varphi-\zeta_{\DM})=\frac{\delta_\varphi}{1+w_s}-\delta_{\DM} \, .
\label{eq:Sdec}
\end{eqnarray}
Using Einstein's eqs. (\ref{Einstein perturbed}) one finds,
\begin{eqnarray}
\tilde{\delta}_\DE=\frac{\rho_\DM (1+w_\DE)}{\rho_\DM + \rho_\DE (1+w_\DE)} \Big[{\cal S}
\;+\;
 3 \cH \frac{ \Delta\theta}{k^2} \Big] \label{Qexplicit} \, .
\end{eqnarray}

Because $\zeta_{\DM}$ is  conserved ${\cal S}'=3 \zeta_\varphi'$. From eq.~\eqref{zetadot} and eq.~\eqref{Qexplicit} it follows ,
\begin{align}
\left\{
\begin{array}{l}
{\cal S}' = 
 -\,{\cal S}\,\dfrac{3(c_{a,\varphi}^{2}-1)\,\cH\,\rho_{\DM}}
          {\rho_{\DM} + (1+w_{\varphi})\,\rho_{\varphi}}
 - \Delta\theta\,\dfrac{9\,(c_{a,\varphi}^{2}-1)\,\cH^{2}\,\rho_{\DM}}
          {k^{2}\,\big(\rho_{\DM} + (1+w_{\varphi})\,\rho_{\varphi}\big)} \, , \\[2ex]
\Delta\theta' = 
 -\cH\,\Delta\theta
 + {\cal S}\,\dfrac{k^{2}\,\rho_{\DM}}
        {\rho_{\DM} + (1+w_{\varphi})\,\rho_{\varphi}}
 + \Delta\theta\,\dfrac{3\cH\,\rho_{\DM}}
        {\rho_{\DM} + (1+w_{\varphi})\,\rho_{\varphi}}\, .
\end{array}
\label{system-entropy}
\right.
\end{align}

Since no source terms appear, if the initial condition satisfy $S_i=\Delta \theta_i=0$
this condition will be preserved outside the horizon. This corresponds to the adiabatic solution
where curvature perturbation $\zeta_{\rm tot}\equiv  -\cH \delta \rho_{\rm tot}/\rho_{\rm tot}'-\Phi$ is constant outside the horizon.

\subsection{Multi-Fluid with Energy Exchange}
We now turn to the non trivial case where the fluids are not isolated. 
We again subtract  eqs.~\eqref{eq:DEpert2} and \eqref{Continuity pert DM 2} to obtain
\begin{eqnarray}
\Delta\theta'&\equiv &  -(\cH+\lm)\Delta\theta+\frac{k^2}{1+w_\DE}\tilde{\delta}_\DE \, ,
\label{deltatheta_interacting}
\end{eqnarray}
where  
\begin{eqnarray}
\tilde{\delta}_\DE \equiv \delta_\DE + \Big[ 3\cH (1+w_\DE) + R_\DM \lm \Big] \frac{\theta_\DE}{k^2} \, .
\end{eqnarray}
The adiabatic initial condition corresponds to $\Delta\theta=0$ and $\tilde{\delta}_\DE=0$.
The isocurvature perturbation is now,
\begin{eqnarray}
{\cal S}\equiv 3(\zeta_\varphi-\zeta_{\DM})=3\cH\left(\frac{\delta\rho_\varphi}{\rho_\varphi'}-\frac{\delta\rho_{\DM}}{\rho'_{\DM}} \right) \, .
\label{Sdef}
\end{eqnarray}
that differs from (\ref{eq:Sdec}) due to the energy exchange.

From Einstein's eqs. (\ref{Einstein perturbed}) one finds,
\begin{eqnarray}
\tilde{\delta}_\DE = R_\DM \frac{3\cH \rho_\DE(1+w_\DE) + \lm \rho_\DM}{3 \cH \left( \rho_\DM + \rho_\DE (1+w_\DE)\right)} \Big[ \frac{ 3\cH - {\lm} }
{3 \cH}{\cal S}
\;+\;
3 \cH \frac{\Delta \theta}{k^2} \Big] \, ,
\end{eqnarray}
with $R_\DM = \rho_\DM / \rho_\DE$.

The evolution of $\Delta \theta$ and $S$ is in this case determined by
\begin{align}
\left\{
\begin{array}{l}
{\cal S}' = A(\eta) S+B(\eta) \Delta\theta \, , \\[2ex]
\Delta\theta' = C(\eta) S+D(\eta) \Delta\theta \, .
 \end{array}
\label{system-entropy-interacting}
\right.
\end{align}
where
\begin{equation}\notag
\begin{aligned}
& C(\eta) = \frac{R_\DM k^2 (3\cH - \lm) (3\cH (1+w_\DE) + R_\DM \lm)}{9 \cH^2 (1+w_\DE)(1+w_\DE + R_\DM)}\, ,\\
&D(\eta) = -\frac{\cH \left[ 1+w_\DE - 2R_\DM \right] + \lm \left[ 1+w_\DE + R_\DM - \frac{R_\DM}{1+w_\DE} \right]}{1+w_\DE + R_\chi} \, ,\\
\end{aligned}
\end{equation}

\begin{equation}\label{eq:coefficient interacting}
\begin{aligned}
& A(\eta) = \frac{1}{2 \rho_{\rm tot} \left[ 1+w_\DE + R_\DM \right] (3 \cH - \lm) \left[ 3\cH (1+w_\DE) + R_\DM \lm \right]} \Big\{ 54 \cH^3 R_\DM \rho_{\rm tot}(1+w_\DE) (c_{a,\DE}^{2}-1) + \\
& \hspace{4cm} + 3 \cH^2 \lm \rho_\DE \Big[ (1+w_\DE)^2 (1+3 w_\DE)+ (1+w_\DE) (13-12 c_{a,\DE}^{2} + w_\DE) R_\DM + \\
& \hspace{4cm} + (5+9w_\DE-6 c_{a,\DE}^{2}(1+2w_\DE)) R_{\DM}^{2} + (6 c_{a,\DE}^{2}- 7 )R_{\DM}^{3} \Big]\\
& \hspace{4cm} + 2 R_\DM \rho_{\rm tot} \lm \Big[ \Big( 1+w_\DE + (1+c_{a,\DE}^{2})R_\DM + \frac{R_{\DM}^{2}}{1+w_\DE} \Big) \lm^2 + 2(1+w_\DE + R_\DM) \lm' \Big] + \\
&\hspace{4cm} + 2\cH \rho_\DM \Big[ (1+w_\DE)\Big( 3 (1+w_\DE) \lm' + (3c_{a,\DE}^{2}+3_\DE -5) \lm^2 \Big) + 3 (1+w_\DE)^2 \frac{\lm'}{R_\DM} + \\
&\hspace{4cm} + R_\DM \Big( (3c_{a,\DE}^{2} (w_\DE-1) -2 (2+w_\DE)) \lm^2 - 3 \lm' \Big) + \\
&\hspace{4cm} + R_{\DM}^{2} \Big( (w_\DE -2 -6c_{a,\DE}^{2}(1+w_\DE))\lm^2 - 3(1+w_\DE) \lm' \big)  - 3 R_{\DM}^{3} \lm^2 \Big] \Big\} \, ,\\
& B(\eta) = - \frac{9 \cH^2 R_\DM (1+w_\DE)}{k^2 (3\cH - \lm) \left[ (1+w_\DE) + R_\DM \right] \left[ 3\cH (1+w_\DE) + R_\DM \lm \right]} \Big[ \left( \lm' + 9 (1-c_{a,\DE}^{2})\cH^2+ (3c_{a,\DE}^{2}-1)\cH \lm \right) + \\
& \hspace{4cm} +  \frac{R_{\DM}}{1+w_\DE} \left( \lm'+c_{a,\DE}^{2}\lm^2 + (5-3c_{a,\DE}^{2}) \cH \lm \right) + \frac{R_{\DM}^{2}}{(1+w_\DE)^2} \lm^2 \Big]\, .
\end{aligned}
\end{equation}
This shows that the adiabatic initial condition $S=\Delta \theta=0$ is preserved outside the horizon. The stability of such system needs also to be studied, but we postpone this to future work. This system also allows to study the evolution of non adiabatic initial conditions.

\section{Fluid Description of a Scalar Field}\label{app 1}

In this appendix we review the correspondence between field and fluid equations that was used in several parts of the paper, see \cite{mainini2008scalar} for further details.

The gravitationally minimally coupled Lagrangian of a scalar field is given by\\
\be\label{scalar field lagrangian}
\mathcal{L}_\varphi = -\sqrt{-g} \left[ g^{\mu \nu}\partial_\mu \varphi \partial_\nu \varphi + V(\varphi) \right]\, ,
\ee
with an associated stress-energy tensor
\be\label{general stress-energy tensor of scalar field}
T_{\mu \nu}^{\varphi} \equiv -\frac{2}{\sqrt{-g}} \frac{\delta \mathcal{L}_\varphi}{\delta g^{\mu \nu}}= \partial_\mu \varphi \partial_\nu \varphi -\frac 1 2 \eta_{\mu\nu} (\partial \varphi)^2 \; .
\ee
Expanding \eqref{general stress-energy tensor of scalar field} around the background, we get the following energy momentum tensor\\
\be\label{matrix tensor of scalar field}
T^{\mu}_{\DE, \nu} = -\gridmatrix{ \frac{\bvarphi'^2}{2 a^2} + V + \frac{\bvarphi'}{a^2} \delta \bvarphi' + \partial_\varphi V \delta \varphi - \frac{\bvarphi'^2}{a^2}\Psi}{-\frac{\bvarphi'}{a^2} \partial^i \delta \bvarphi }{\frac{\bvarphi'}{a^2} \partial_i \delta \bvarphi }{\delta^{i}_{j} \left[ V - \frac{\bvarphi'^2}{2 a^2} - \frac{\bvarphi'}{a^2}\delta \bvarphi' +\partial_\varphi V \delta \varphi + \frac{\bvarphi'^2}{a^2} \Psi \right]} + \mathcal{O}(\delta \varphi, \Psi, \Phi) \, .
\ee
On the other hand a general fluid, non necessarily perfect or adiabatic, is characterized by an energy momentum tensor,\\
\be\label{def fluid}
T^\mu_\nu \equiv (\rho_\varphi + p_\varphi) u^{\mu}_{\varphi} u_{\nu, \varphi} + p_\varphi g_{\mu \nu} +\Sigma^\mu_{\varphi ,\nu} \; ,
\ee
where the 4-velocity $u^\mu$ and anisotropic stress $\Sigma^{\mu\nu}$ satisfy,
\begin{equation}
g_{\mu \nu} u^\mu u^\nu = -1\,,~~~~\partial_\mu\Sigma^\mu_{\varphi,\nu}=0\, ,~~~~~  \Sigma^\mu_{\varphi, \mu} = 0\; .
\end{equation}

Expanding in small perturbations around the ground state,
\begin{equation}
u^\mu \approx \frac{1}{a} \left( 1 - \Psi  \, , \, v^i \right) \, ,
\end{equation}
one finds,
\be\label{matrix tensor fluid}
T^\mu_\nu = \gridmatrix{-\brho_\varphi (1+\delta_\varphi)}{-\brho_\varphi(1 + w_\varphi) v_{\varphi}^{i} }{\brho_\varphi(1 + w_\varphi) v_{\varphi,i}}{\delta^{i}_{j}\left[ \brho_\varphi w_\varphi + \delta p_\varphi \right] + \Sigma^i_j} + \mathcal{O}(\delta_\varphi, \delta p_\varphi, v_{\DE}^{i}, \Sigma^i_j) \, .
\ee
At the background level one finds,
\be\label{background dictionary}
\begin{split}
&\bvarphi'^2 = a^2 \brho_\varphi(1+w_\varphi)\, ,\\
&w_\varphi = \frac{\bvarphi'^2-2a^2V}{\bvarphi'^2+2a^2V}\, .
\end{split}
\ee
At linear order in perturbations the correspondence reads,\\
\be\label{linear dictionary}
\begin{split}
&\delta_\varphi = (1+w_\varphi)\left[ \frac{\delta \varphi'}{\bvarphi'} + \frac{a^2 \partial_\varphi V}{\bvarphi'} \frac{\delta \varphi}{\bvarphi'}- \Psi\right]\, ,\\
&\theta_\varphi \equiv \partial_i v^{i}_{\varphi} = -\frac{1}{\bvarphi'}\nabla^2 \delta \varphi \, ,\\
&\delta p_\varphi = \brho_\varphi (1+w_\varphi)\left[ \frac{\delta \varphi'}{\bvarphi'} - \frac{a^2 \partial_\varphi V}{\bvarphi'} \frac{\delta \varphi}{\bvarphi'}-\Psi\right] \, ,\\
&\Sigma^{i}_{\varphi,j} = 0\; .
\end{split}
\ee
Note that the pressure has a contribution that depends on the scalar potential. As a consequence setting to zero the density perturbations of the scalar is not equivalent to freezing the scalar field perturbations.\\

Importantly for a scalar field the pressure fluctuation cannot be written as a function of the density contrast. This means that while to linear order the scalar field is a perfect fluid in the sense that the anisotropic stress vanishes, it is not barotropic. As consequence the pressure, given by eq. (\ref{pressure ad+nad}) is in general different from the adiabatic pressure. Indeed,
\be\label{speed of sound}
c_{s,\varphi}^{2} \equiv \frac{\delta p_\varphi}{\delta \rho_\varphi} \Big|_{\theta_\varphi = 0} = 1 \, .
\ee
This is at the heart of the physics. A rolling scalar field has sound speed 1 in the rest frame that prevents DE from clustering. The notion of adiabatic velocity $c_a^2=p'/\rho'$ becomes useful outside the horizon because for adiabatic initial conditions the pressure becomes indeed adiabatic.

\pagestyle{plain}
\bibliographystyle{jhep}
\small
\bibliography{biblio}

\end{document}